\documentclass[journal,10pt]{IEEEtran}

\usepackage{color,array,amsthm}
\usepackage{graphicx}
\usepackage{amsmath}
\usepackage{amssymb}
\usepackage{stfloats}
\usepackage{amsthm}
\usepackage{amsfonts}
\usepackage{color}
\usepackage{graphicx}
\usepackage{subfigure}
\usepackage{ragged2e}

\begin{document}

\title{A Hardware Prototype of Wideband High-Dynamic Range ADC}

\author{Satish Mulleti,~\IEEEmembership{Member,~IEEE}, Eliya Reznitskiy, Shlomi Savariego, Moshe Namer, Nimrod Glazer, ~\IEEEmembership{Member,~IEEE}, Yonina C. Eldar,~\IEEEmembership{Fellow,~IEEE}

\thanks{S. Mulleti is with the Department of Electrical Engineering, Indian Institute of Technology (IIT) Bombay, Mumbai, India; E. Reznitskiy, S. Savariego, N. Glazer, and Y. C. Eldar is with the Faculty of Math and Computer Science, Weizmann Institute of Science, Rehovot, Israel. Emails: mulleti.satish@gmail.com, yonina.eldar@weizmann.ac.il}
\thanks{This research was partially supported by a research grant from the Estate of Tully and Michele Plesser, from the European Research Council (ERC) under the European Union’s Horizon 2020 research and innovation program (grant No 101000967) by the Israel Science Foundation under grant no. 0100101.}
\thanks{Manuscript received XXX, XX, 2023; revised XXX, XX, 2023.}}

\markboth{IEEE Transactions on Vehicular Technology,~Vol.~XX, No.~XX, XXX~2023}
{}

\maketitle




\begin{abstract}
Key parameters of analog-to-digital converters (ADCs) are their sampling rate and dynamic range. Power consumption and cost of an ADC are directly proportional to the sampling rate; hence, it is desirable to keep it as low as possible. The dynamic range of an ADC also plays an important role, and ideally, it should be greater than the signal's; otherwise, the signal will be clipped. To avoid clipping, modulo folding can be used before sampling, followed by an unfolding algorithm to recover the true signal. In this paper, we present a modulo hardware prototype that can be used before sampling to avoid clipping. Our modulo hardware operates prior to the sampling mechanism and can fold higher frequency signals compared to existing hardware. We present a detailed design of the hardware and also address key issues that arise during implementation. In terms of applications, we show the reconstruction of finite-rate-of-innovation signals which are beyond the dynamic range of the ADC. Our system operates at six times below the Nyquist rate of the signal and can accommodate eight-times larger signals than the ADC's dynamic range. 
\end{abstract}


\section{INTRODUCTION}
Analog-to-digital converters (ADCs) bridge real-world analog signals and digital processors on which signals can be processed efficiently. Typically, ADCs measure instantaneous uniform samples of analog signals to represent them digitally. A key parameter in such conversion is the sampling rate. Power consumption and cost of an ADC increase with the increase in the sampling rate. Hence, keeping the sampling rate as low as possible is desirable. Theoretically, the sampling rate has to be greater than the Nyquist rate for perfect reconstruction of bandlimited signals. Apart from the sampling rate, there are several other aspects of an ADC which play a key role in faithful sampling and reconstruction, especially when the sampling frameworks are implemented in hardware.

The dynamic range of an ADC plays a crucial role in sampling an analog signal. Generally, ADC's dynamic range should be larger than the signal's; otherwise, the signal gets clipped. A few approaches exist to recover the true samples from clipped ones for bandlimited signals \cite{marks_clipping2, marks_clipping1}. These approaches rely on the correlation among the samples when they are measured at a very high rate compared to the Nyquist rate. The requirement of a high sampling rate is a drawback of these approaches. 

Several preprocessing approaches to avoid clipping exist, such as automatic gain control (AGC) \cite{perez2011automatic,mercy1981review}, companding \cite{landau_compander,landau_distorted_bl}, and modulo folding \cite{uls_tsp,uls_romonov,bhandari2021unlimited,eyar_moduloicassp, azar2022robust}. Among these, modulo folding is the most recent approach that need not be differentiable like companding and does not suffer from stability issues of the feedback amplifiers used in AGCs. In the modulo framework, the signal is folded to lie within the ADC's dynamic range, and then the folded signal is sampled using a conventional ADC. Theoretical guarantees for recovering bandlimited signals from folded samples are presented in \cite{uls_tsp}. The results state that a bandlimited signal can be uniquely recovered from its folded samples provided that they are sampled above the Nyquist rate \cite{uls_tsp}.

Several algorithms for \emph{unfolding} or recovering the true samples of a bandlimited signal from a modulo or folded samples are presented in \cite{uls_tsp,uls_romonov, eyar_moduloicassp}. These unfolding algorithms can be compared in terms of sampling rate, amount of unfolding they can handle, and noise robustness. The algorithm proposed in \cite{uls_tsp} requires almost 17 times higher sampling rate than the Nyquist rate. The approaches in \cite{uls_romonov} and \cite{eyar_moduloicassp,azar2022robust} operate at relatively lower sampling rates but require the knowledge of ADC's dynamic range. In contrast, the method proposed in \cite{eyar_moduloicassp,azar2022robust} requires a lower sampling rate, even in the presence of noise, compared to the algorithms presented in \cite{uls_tsp,uls_romonov}. Modulo sampling is also extended to different problems and signal models such as periodic bandlimited signals \cite{bhandari2021unlimited}, finite-rate-of-innovation (FRI) signals \cite{uls_fri}, sparse vector recovery \cite{uls_sparsevec}, direction of arrival estimation \cite{uls_doa}, computed tomography \cite{uls_radon}, and graph signals \cite{uls_graph}.

Beyond theoretical works, there also exist a few related hardware prototypes. High-dynamic-range ADCs, also known as \emph{self-reset} ADCs, are discussed in the context of imaging \cite{sradc_park, sradc_sasagwa, sradc_yuan, krishna2019unlimited}. These hardware architectures measure additional information, such as the amount of folding for each sample or the sign of the folding together with the folded samples. The additional information might enable simpler recovery at the expense of complex circuitry. Importantly, additional bits are required during the quantization process to store or transmit the side information. 

Krishna et al. presented a hardware prototype that encodes the side information by using two bits \cite{krishna2019unlimited}. The architecture is designed to record the sign of the slope of the signal at each sample that lies outside the ADC's dynamic range. In a conventional ADC, a sample and hold (S/H) circuit is used to hold the sampled value for a prescribed period of time, during which quantization is performed on the sample. A folding circuit is used after S/H  to realize the modulo sampling \cite{krishna2019unlimited}. In this architecture, the S/H circuit has to hold the sampled value for folding and quantization, resulting in a larger holding time than a conventional ADC. A large holding time results in slower ADCs, which may not be helpful in applications with high-frequency signals. The resulting hardware circuit is able to fold signals up to 300 Hz, where the signal's amplitude should be less than three times the ADC's dynamic range.


Modulo hardware prototypes where the modulo part is implemented prior to the sampler are presented in \cite{bhandari2021unlimited,Bhandari:2022:J,mod_hyst}.
In these works, the authors are focused on different signal models, hardware limitations, and algorithms rather than providing details of the hardware circuitry. It was shown that the modulo hardware is able to fold low-frequency ($<300$ Hz) signals that are tenfold larger than the ADC's dynamic range. However, it is not clear how the hardware performs for high-frequency signals, and many details of the circuitry are omitted.

In practical applications, the frequency range of the signals can vary from a few kHz to several MHz. For example, the finite-rate-of-innovation (FRI) model is widely used to represent signals in time-of-flight applications such as ultrasound, sonar, and radar \cite{vetterli,eldar_sos}. These FRI signals have frequencies much higher than 300 Hz, and hence current hardware prototypes can be used, especially when the signal's bandwidth ranges up to a few kHz. Hence it is desirable to design and develop a modulo sampler that can operate at high frequencies while folding signals faithfully.

In this paper, we present a modulo hardware prototype that can be used for modulo sampling of signals up to 10 kHz. We show that by using our algorithm \cite{eyar_moduloicassp}, it is able to reconstruct bandlimited and FRI signals faithfully. 
In the following, we present the contributions and the features of the proposed hardware system. 
\begin{itemize}
    \item We design our hardware components to be able to fold signals up to 10kHz. Existing hardware shows results for signals below 300 Hz.
   
    \item The hardware prototype is designed to perform folding prior to the sampler, unlike the hardware in \cite{krishna2019unlimited}, which operates in the hold part of the sampler.  Thus, the suggested system can utilize faster ADCs with shorter hold times.
    
    \item In the proposed hardware prototype, modulo folding is realized through a feedback mechanism. At the time instants when the input signal goes beyond the ADC's dynamic range, a trigger signal is generated by using comparators. The trigger then activates a direct voltage generator that adds to the input signal to bring it within the dynamic range. This mechanism imposes a delay between the trigger time and the folding instance. We address this key issue of the hardware, which is not considered in previous works. 
    By using the signal's smoothness and the feedback loop's time delay, 
    we propose a hardware solution to avoid clipping that occurs due to the delay issue. 
    

    \item The designed hardware prototype can operate at a maximum voltage of 11.75 v. The limitation is largely due to the use of a 15v subtractor or adder in the feedback loop, which enables a fast slew rate in the transitions of $\pm 2\lambda$. At high frequencies, these components can not be used at voltage above 15 v. In addition, we used an ADC with a dynamic range $[-1.25, 1.25]$. Hence, the hardware can fold signals which are eight times larger than the dynamic range of the signal.

    
    \item For demonstration, we consider sampling and reconstruction of bandlimited and FRI signals. For FRI signals, we use a lowpass sampling kernel prior to modulo folding. The filter removes unwanted information in the signal and allows sub-Nyquist sampling. Using our algorithm presented in \cite{eyar_moduloicassp,azar2022robust}, we show reconstruction of bandlimited signals from their folded samples measured through the hardware. We show that the combination of the proposed hardware and low-rate algorithm is able to reconstruct the signals by using a low-dynamic range ADC. In particular, for FRI signals, we show that the FRI parameters can be estimated with sub-Nyquist samples by utilizing the fact that our algorithm operates at the lowest possible rate. 
    
    
\end{itemize}


The paper is organized as follows. In the next section, we discuss the signal model considered and the sampling and reconstruction framework in the presence of modulo hardware. In Section III, we present the hardware system by explaining its working principle and discussing the components of the system. In Section IV, we show the hardware's signal folding and reconstruction abilities.

\section{Signal Model and System Description}
\begin{figure*}
	\centering
	\includegraphics[width=4 in]{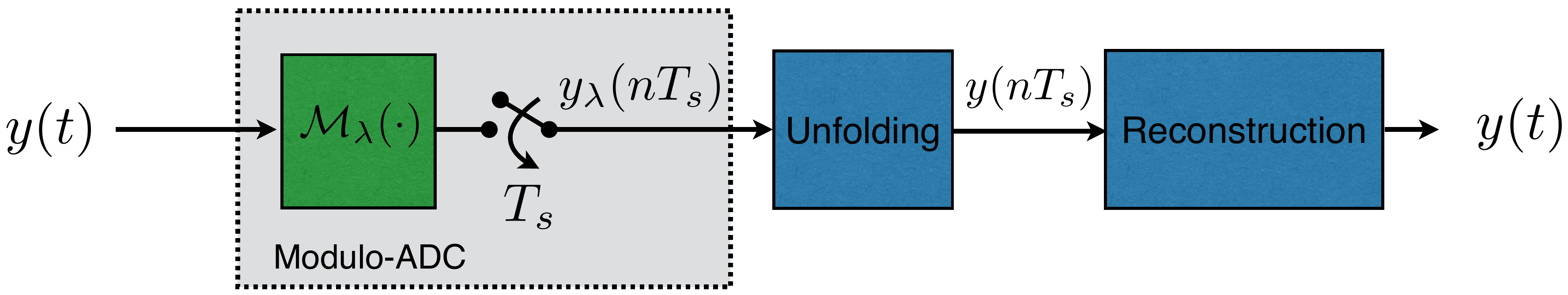}
	\caption{A schematic of modulo-sampling and reconstruction of bandlimited signals.}
	\label{fig:bl_framework}
\end{figure*}
In this section, we consider modulo sampling for signals whose amplitudes lie beyond the bandwidth of the ADC's dynamic range. The class of signals that can be folded by modulo hardware can be very large; however, the recovery is limited by existing unfolding algorithms. For example, most unfolding algorithms are designed for bandlimited signals. Given this, we consider bandlimited signals as input to the modulo hardware and corresponding unfolding algorithms. We use a lowpass sampling kernel for FRI signals to make them bandlimited and, at the same time, reduce the sampling rate following the sub-Nyquist framework \cite{eldar_sos, mulleti_kernal}. 

Consider a $\omega_c$-bandlimited signal $y(t)$ such that its Fourier transform $Y(\omega)$ vanishes outside the frequency interval $[-\omega_c, \omega_c]$. The signal can be perfectly reconstructed from its uniform samples measured at the Nyquist rate $\omega_{Nyq} = 2\omega_c$ rad/sec. provided that the ADC's dynamic range is above the signal's dynamic range. Specifically, if the dynamic range of the ADC is $[-\lambda, \lambda]$ for some $\lambda >0$ then it is assumed that $|y(t)|\leq \lambda$ for perfect reconstruction. If $|y(t)|>\lambda$, then the signal and its samples will be clipped, and perfect reconstruction is not guaranteed. In the latter scenario where $|y(t)|>\lambda$, one can either increase the dynamic range of the ADC or use prepossessing to avoid clipping. We consider the later solution where the modulo operation $\mathcal{M}_{\lambda}(\cdot)$ is applied to the signal $y(t)$ to restrict its dynamic range to $[-\lambda, \lambda]$. The output of the modulo operator in response to input $y(t)$ is given as
\begin{equation}
 y_\lambda(t) =	\mathcal{M}_{\lambda}(y(t)) = (y(t)+\lambda)\,\, \text{mod}\,\, 2\lambda -\lambda.	
	\label{eq:mod_opp}
\end{equation}
The folded signal $y_\lambda(t)$ is then sampled to get discrete measurements $y_\lambda(nT_s)$. Due to modulo folding, $y_\lambda(t)$ is no longer bandlimited. To recover $y(t)$ while sampling slightly above the Nyquist rate of the input, one first applies an unfolding algorithm to recover $y(nT_s)$ from $y_\lambda(nT_s)$ \cite{eyar_moduloicassp,azar2022robust,mulleti2022modulo}. Then $y(t)$ is reconstructed from $y(nT_s)$ by assuming that the sampling is performed above the Nyquist rate.

A schematic of modulo sampling and reconstruction framework is shown in Fig.~\ref{fig:bl_framework}. It consists of a modulo-ADC followed by unfolding and reconstruction blocks. The modulo-ADC is comprised of a modulo-folding block followed by a conventional uniform sampler. The unfolding operation is implemented in the digital domain, and it should operate at the lowest possible sampling rate. To this end, we use the $B^2R^2$ algorithm for unfolding \cite{eyar_moduloicassp,azar2022robust}, which samples efficiently compared to other algorithms for bandlimited signals. Low-rate sampling and low-dynamic range requirements significantly reduce the power consumption and cost of the ADC.

Our objective is to demonstrate a robust hardware prototype of modulo ADC as discussed next.

\begin{figure}
	\centering
	\includegraphics[width= 3 in]{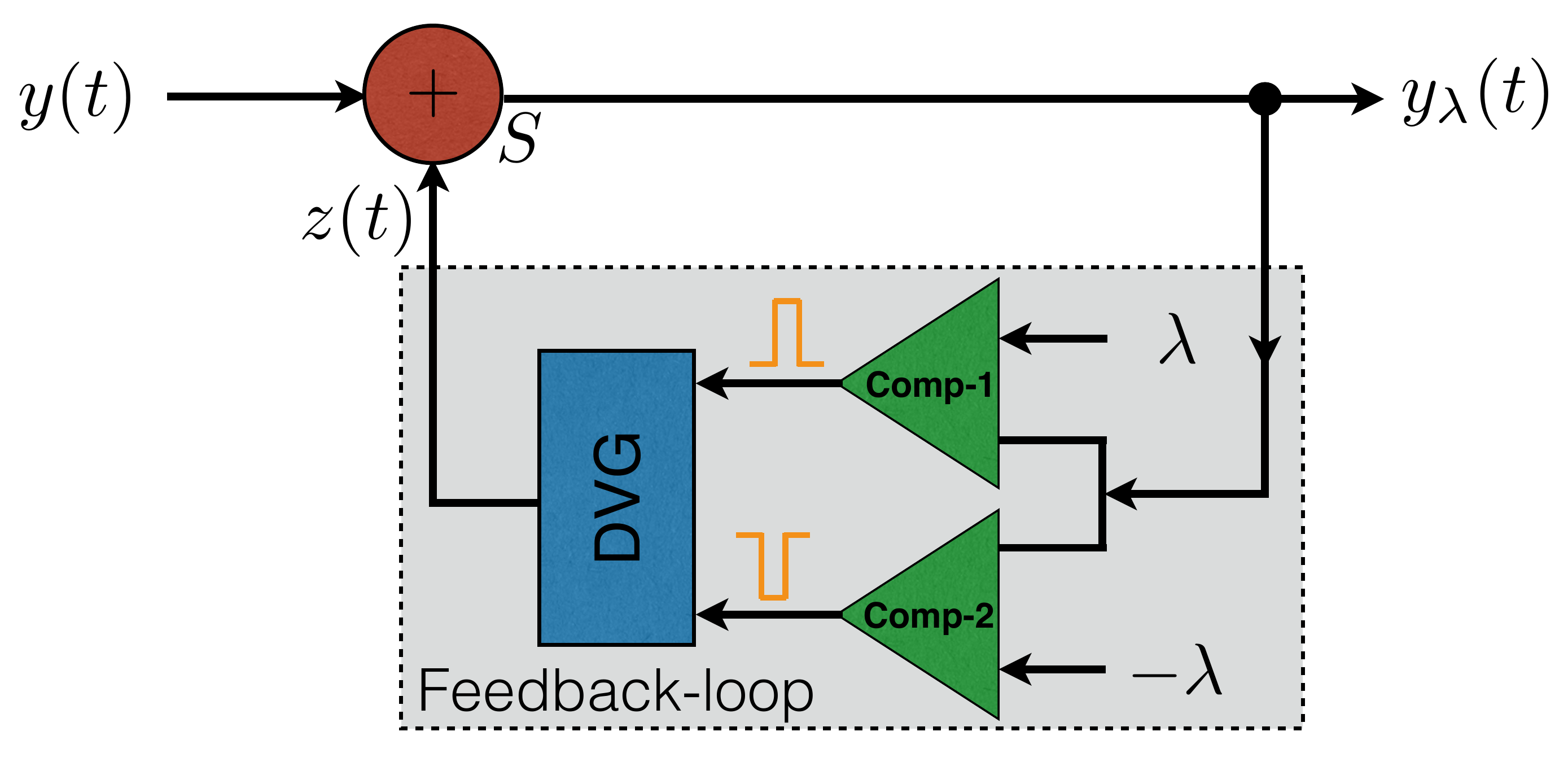}
	\caption{Folding principle.}
	\label{Fig:folding}
\end{figure}
\section{Modulo Hardware Prototype}
In this section, we discuss the prototype of our modulo hardware. The modulo block's working principle and design will be discussed first, followed by its hardware implementation.
\subsection{Working Principle of Modulo Block}
The principle of computing $y_\lambda(t)$ from $y(t)$ is shown by the block diagram in Fig.~\ref{Fig:folding}. The system comprises an adder $S$, a direct-voltage generator (DVG), and two comparators, Comp-1 and Comp-2. To understand the working flow, let us first assume that for some time instant $t_1$, we have that $|y(t)|<\lambda$ for all $t<t_1$. Hence $y_\lambda(t) = y(t)$ and $z(t) = 0$ for all $t<t_1$. At $t=t_1$, let $|y(t)|$ cross $\lambda$. If $y(t_1)>\lambda$, then Comp-1 triggers a positive value. Else if, $y(t_1)<\lambda$, Comp-2 triggers a negative value. The DVG is designed such that for each positive input value, its output signal level increases by $-2\lambda$, whereas, for a negative input value, it decreases its output voltage by $2\lambda$. Hence, in the current example, DVG generates a signal $z(t) = \text{sgn}(y_\lambda(t_1)) 2\lambda u(t-t_1)$ where $u(t)$ is the unit-step function. In this way, by adding or subtracting (using $S$) constant DC signals from $y(t)$ whenever it crosses the dynamic range $[-\lambda, \lambda]$, the amplitude levels of $y_\lambda(t)$ are kept within the ADC's dynamic range.


\begin{figure}[!t]
	\centering
	\includegraphics[width= 3 in]{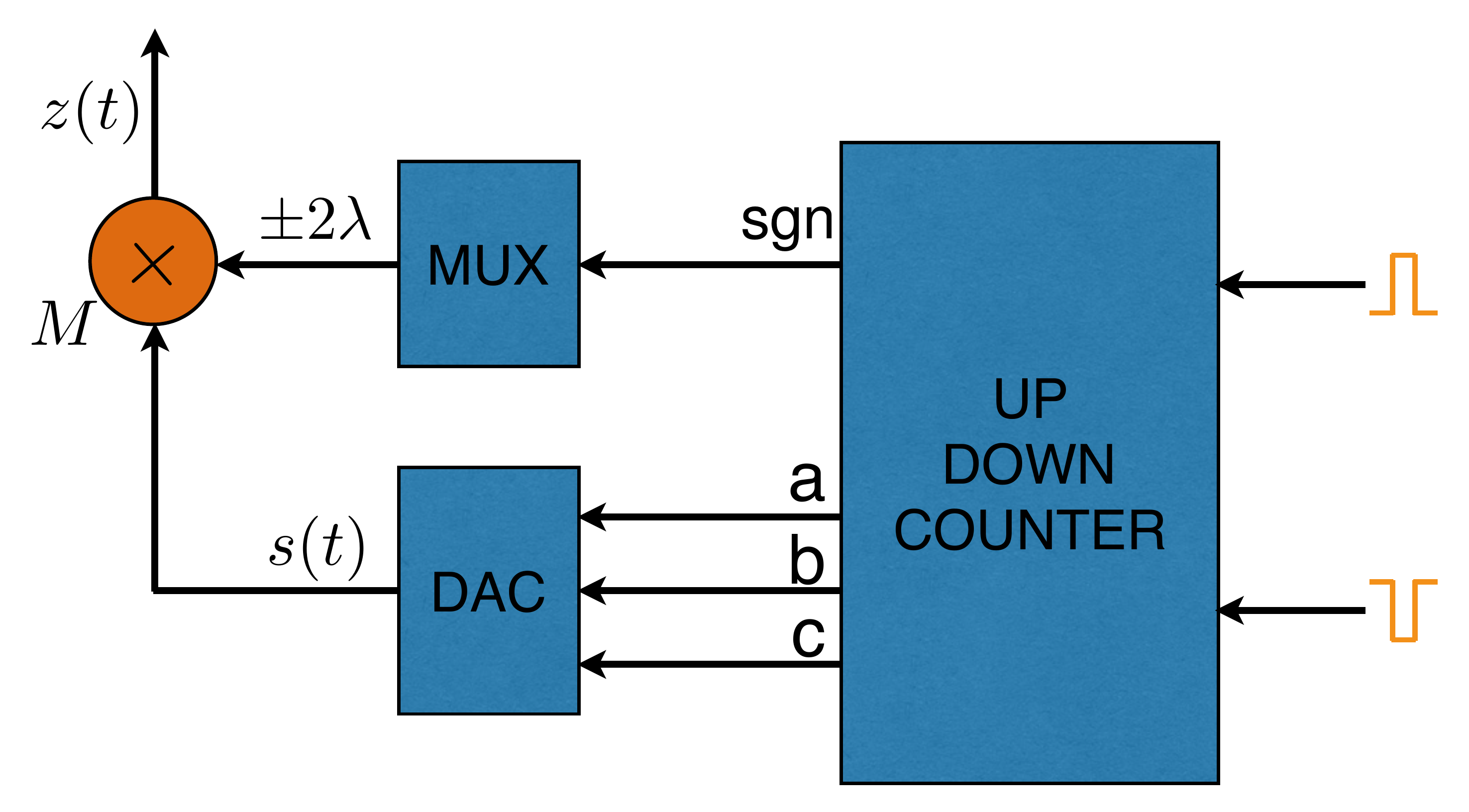}
	\caption{Discrete Voltage Generator (DVG).}
	\label{Fig:dcvg}
\end{figure}

\begin{figure*}[!t]
	\centering
	\includegraphics[width=0.86\textwidth ]{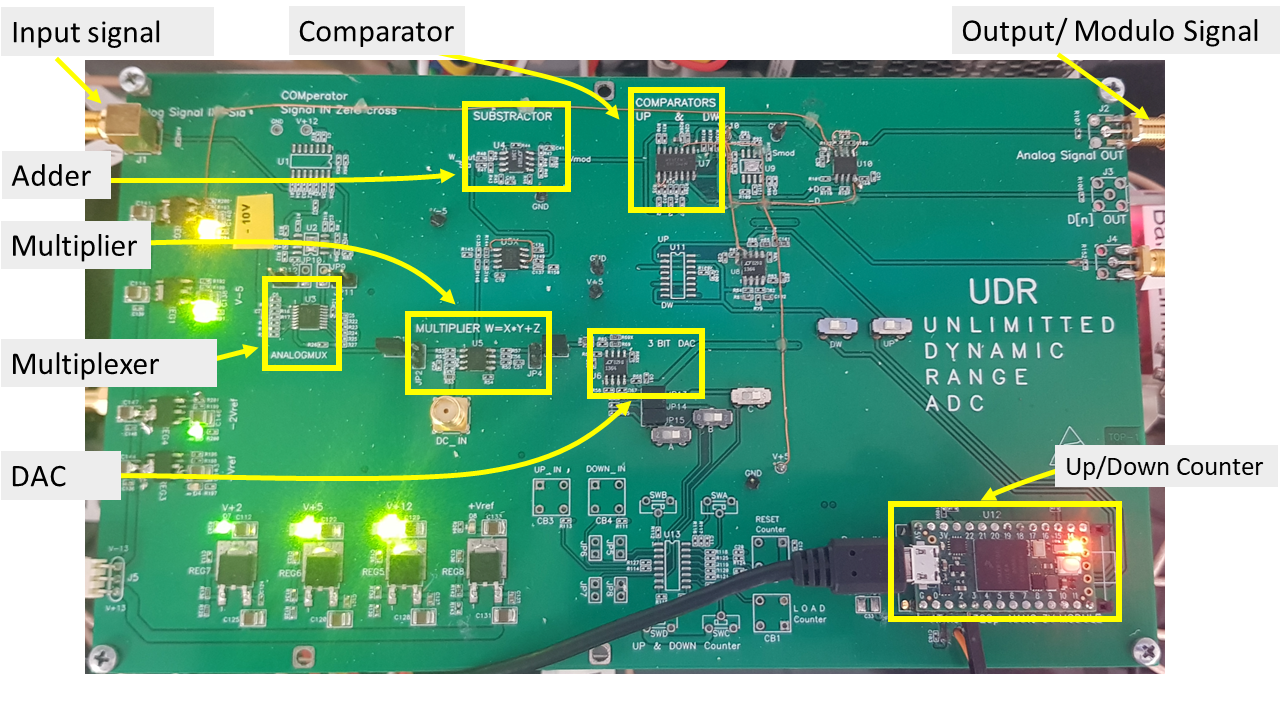}
	\caption{Modulo hardware board.}
	\label{Fig:hw_board}
\end{figure*}

While the parts such as comparators Comp-1 and Comp-2, and adder $S$ can be realized by using off-the-shelf components, DVG is a more involved system due to its feedback nature and requires careful design. Specifically, the feedback loop should follow changes in the input signal in the desired frequency and amplitude ranges. A detailed architecture of DVG is shown in Fig.~\ref{Fig:dcvg}. Its task is to generate a constant voltage signal whose amplitude is a constant multiple of $2\lambda$. Importantly, its output voltage $z(t)$ should increase (or decrease) by $2\lambda$ v for every negative (or positive trigger) at its input. To realize this task in the hardware, we use an up/down counter, a digital-to-analog converter (DAC), a multiplexer (MUX), and a multiplier $M$. 

In the hardware design, we set $\lambda = 1.25$ v.
We start with the DAC, which can generate piecewise constant voltage output $s(t)$ in response to its digital input. Let the resolution, or step size of the DAC, be $\alpha$ v. Then, when the input bits of DAC go from one state to the next, the DAC output increases by $\alpha$ v. On the other hand, when the bits change from the present state to the previous state, output voltage $s(t)$ reduces by $\alpha$ v. Hence, the ADC works in a fashion expected by DVG with the following exceptions: (1) Input to the DAC is bits and one needs to map positive/negative trigger from comparators to these bits; (2) Output of the DAC takes only positive values and are multiple of $\alpha$. A scaling is required to make them multiple of $\pm 2\lambda$. To address the first issue, we employ a UP/DOWN counter whose inputs are the trigger voltages from the capacitors Comp-1 and Comp-2, and the output is bits. For every positive trigger at the input counter, output bits change to the next state, whereas for a negative trigger, they go back to the previous state. By connecting these bits to the input of the DAC, the output of the DAC is controlled by triggers. 

To address the scaling issue, we use a MUX and a voltage multiplier $M$. The MUX and the multiplier are designed, together with a set of amplifiers, such that $s(t)$ is scaled to $z(t)$. A sign bit at the output of the counter, which is a function of the trigger's sign, is used as input to the MUX, which in turn controls the sign of the multiplier's output or $z(t)$'s sign. 

To explain the sequences of events in DVG, let us consider our previous scenario where $|y(t)| <\lambda$ for some $t<t_1$ and at $t_1$, $|y(t)|$ crosses $\lambda$. For $t<t_1$, we have $y(t) = y_\lambda(t)$, $s(t) = 0$, $z(t) = 0$, and all the output bits of the counter are set to be zero. If $y(t_1)> \lambda$, Comp-1 triggers a positive voltage, and the counter's output bits state changes. Specifically, the least significant bit changes to one, and in response, the DAC's output voltage changes to $\alpha$ v. Meanwhile, after the positive trigger, the MUX outputs a voltage $-2\lambda/\alpha$ which is multiplied to $s(t)$ and outputs $z(t)$ as $-2\lambda u(t-t_1)$ as desired.

Next, we discuss the hardware board that realizes the folding operation discussed in this section.

\begin{table}[h!]
\centering
\caption{List of Hardware Components}
\label{table:hardware_components}
\begin{tabular}{||c c c||} 
 \hline
 Component & Model Number & Make \\
 && \\ \hline
 Comparator & LM339 & Texas Instruments \\ 
 && \\
 UP/DOWN Counter & TEENSY4.1 & PJRC \\
 && \\
 Analog MUX & ADG1608 & Analog Devices \\
 && \\
 Analog Multiplier & AD835 & Analog Devices \\
 && \\
 Adder & LT1364 & Analog Devices \\
  && \\ [1ex]
\hline
\end{tabular}
\end{table}

\begin{table}[h!]
		\caption{Up/down counter operation.}
	\centering
	\begin{tabular}{||c c c c c c c||} 
		\hline
		c & b & a & sgn & Counter values & $s(t)$ & $z(t)$  \\ [0.5ex] 
		\hline\hline
		0 & 0 & 0 & 0~(1) & 0~(0) & 0 & 0  \\
		0 & 0 & 1 & 0~(1) & 1~(-1) & 1 & $2\lambda$ ($-2\lambda$)\\
		0 & 1 & 0 & 0~(1) & 2~(-2) & 2 & $4\lambda$ ($-4\lambda$)\\
		0 & 1 & 1 & 0~(1) & 3~(-3) & 3 & $6\lambda$ ($-6\lambda$)\\
		1 & 0 & 0 & 0~(1) & 4~(-3) & 4 & $8\lambda$ ($-8\lambda$)\\
		\hline
	\end{tabular}
	\label{table:counter}
\end{table}

\begin{figure}[!t]
	\centering
      \begin{tabular}{c}
      
	\subfigure[] {\includegraphics[width= 3.2 in]{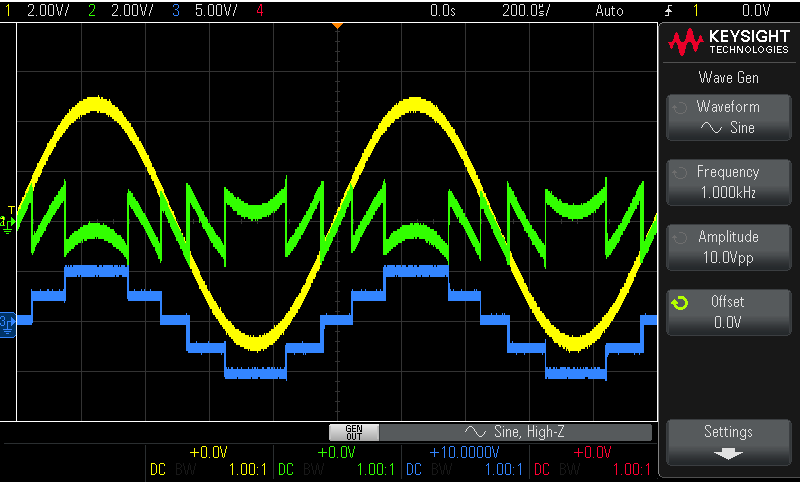}}\label{fig:2fold}\\
 \subfigure[] {\includegraphics[width= 3.2 in]{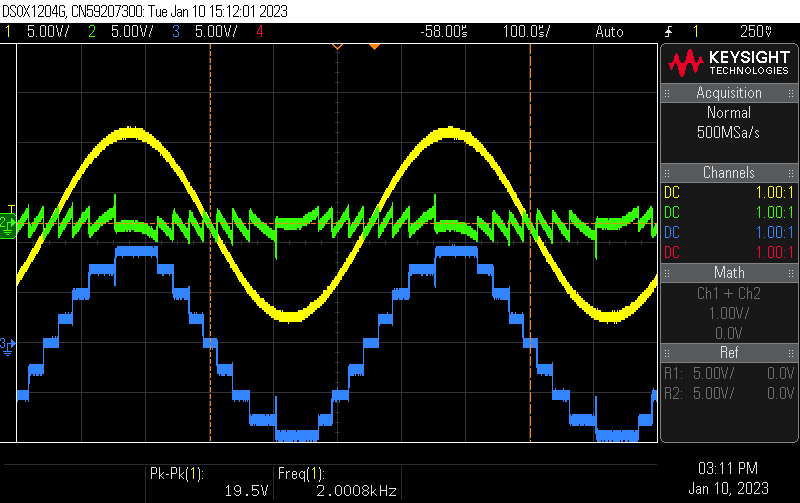}}\label{fig:4fold}
	
 \end{tabular}
 \caption{Screenshots of an oscilloscope capturing input signals (yellow), its folded outputs (green), and the DVG signals (blue): (a) 1kHz sinusoid with maximum amplitude $4\lambda$ and (a) 2kHz sinusoid with maximum amplitude $8\lambda$.}
	\label{Fig:Analysis_of_folding_ability}
\end{figure}

\begin{figure}[!h]
	\centering
	\begin{tabular}{c}
		\subfigure[] {\includegraphics[width=3.2in]{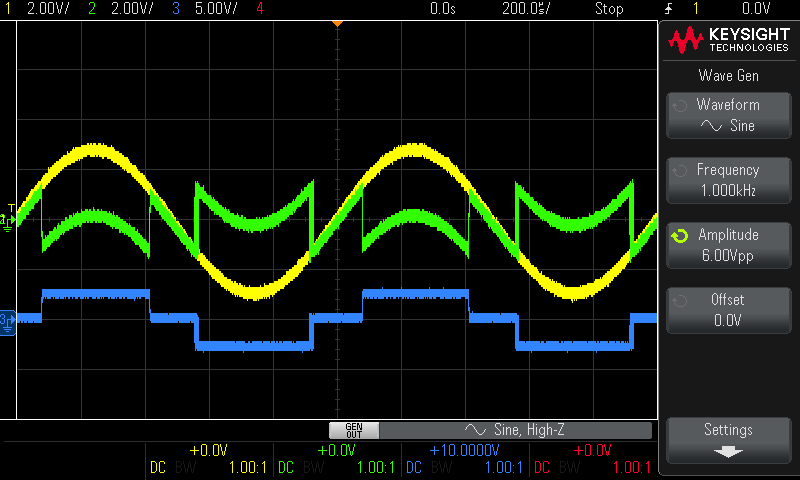}\label{fig:1k_1}} \\
		\subfigure[] {\includegraphics[width=3.2in]{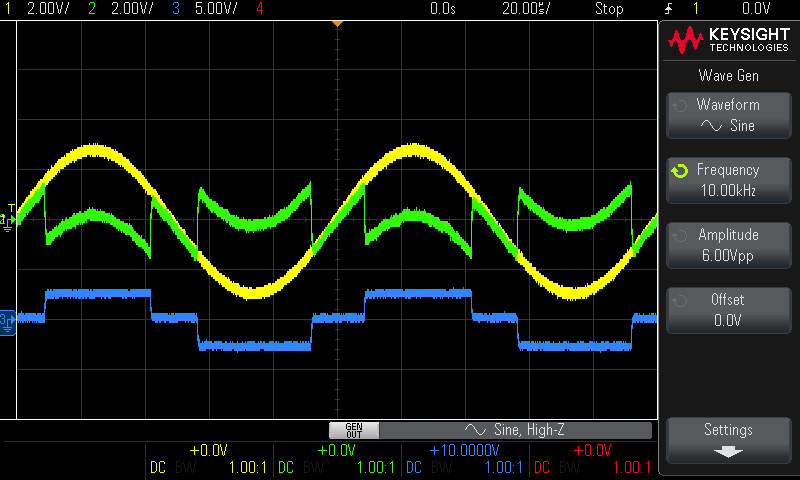}\label{fig:10k_1}} \\
		\subfigure[] {\includegraphics[width=3.2in]{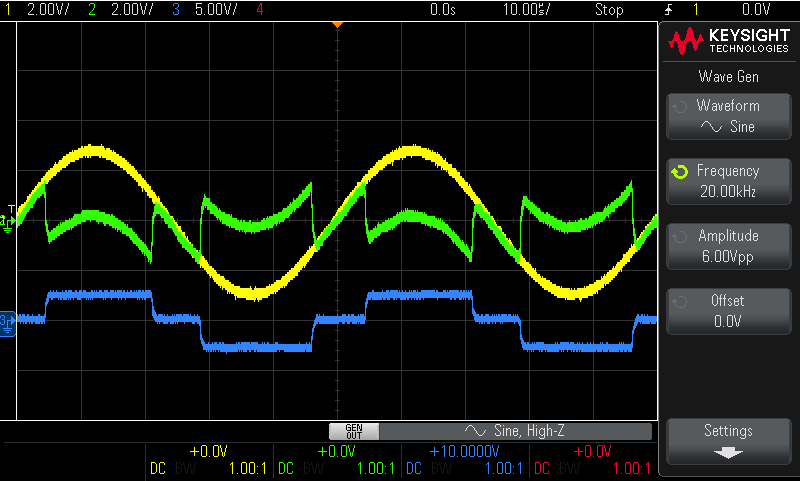}\label{fig:20k_1}} 
	\end{tabular}
	\caption{Analysis of frequency response of the modulo hardware (a)-(c) denote the real-time capture of an oscilloscope capturing the 1kHz, 10kHz, 20kHz, respectively, the sinusoidal input signal (yellow), its folded output (green), the DVG signal (blue).}
	\label{Fig:frequency_response}
\end{figure}

\subsection{Modulo Hardware Board}
Our modulo hardware board is presented in Fig.~\ref{Fig:hw_board}, along with the roles of the major building components. Table~\ref{table:hardware_components} contains a detailed listing of the hardware's components. The board is designed for $\lambda = 1.25$ v. While selecting components for the MUX, multiplier, and amplifiers involved, we observe that these components operate in their linear regions if the operating voltages are less than 12 v. This implies that $|z(t)|\leq 12$ v which limits the maximum value of input signal to $|y(t)| < 9\lambda = 11.75$ v. This is because if $y(t)$ crosses $9\lambda$ then $z(t)$ should be $-10\lambda = 12.5$ v to ensure that $y_\lambda = y(t) + z(t) \in [-\lambda, \lambda]$. Hence, the current design of the hardware can fold and sample signals eight times larger than ADC's dynamic range. This implies that the DVG output should take values from the set $\{0, \pm 2\lambda, \pm 4\lambda, \pm 6\lambda, \pm 8\lambda \}$. This requires the DAC to have five uniform voltage levels at its output (it produces only positive voltages), and a  3-bit DAC and hence a 3-bit counter are used as shown in Fig.~\ref{Fig:dcvg}. Instead of using an off-the-shelf DAC, we build a customized DAC for the hardware. By noting the DAC's output is a linear combination of its three input bits, we used adders LT1364 to realize the DAC. In Table~\ref{table:counter}, we list the values of bits and the counter (denoted as counter values). The three bits (a,b, and c) of the counter are used as input to the DAC, which converts the bits to an analog DC voltage. Here the resolution of the DAC is $\alpha = 1$ v. 




We further analyze the working of the modulo hardware by considering a sinusoidal signal $y(t) = A\, \sin(2\pi f_0 t)$ where $A$ is amplitude, and $f_0$ is the frequency (in Hz). In this experiment, we set $\lambda = 1.25$. First, we analyze the folding ability of the hardware for different amplitude levels. Figure~\ref{Fig:Analysis_of_folding_ability}(a) and (b) depict screenshots of an oscilloscope capturing input $y(t)$ (in yellow), folded output $y_\lambda(t)$ (in green), and the DVG output $z(t)$ (in blue), for two signals with $f_0 = 1$ kHz, $A = 4\lambda$ and $f_0 = 2$ kHz, $A = 8\lambda$, respectively. We observed that the signals are folded back to lie within the dynamic range of the ADC as expected without clipping. 

Next, we discuss the frequency response of the modulo ADC. As in any analog system, the modulo folder's response also depends on the input signal's frequency or bandwidth. In particular, beyond a particular frequency range, components of the hardware and the overall feedback loop may not respond quickly to fast changes in the input signal, as demonstrated in Fig.~\ref{Fig:frequency_response}. We observed that for 1kHz and 10 kHz, the hardware folds the signal accurately. However, for $f_0 = 20$ kHz, folding instants are not symmetric for positive and negative folds. 



In the next section, we discuss several challenges of the modulo hardware and our proposed solutions. 

\begin{figure}[!t]
	\centering
	\includegraphics[width= 3.2 in]{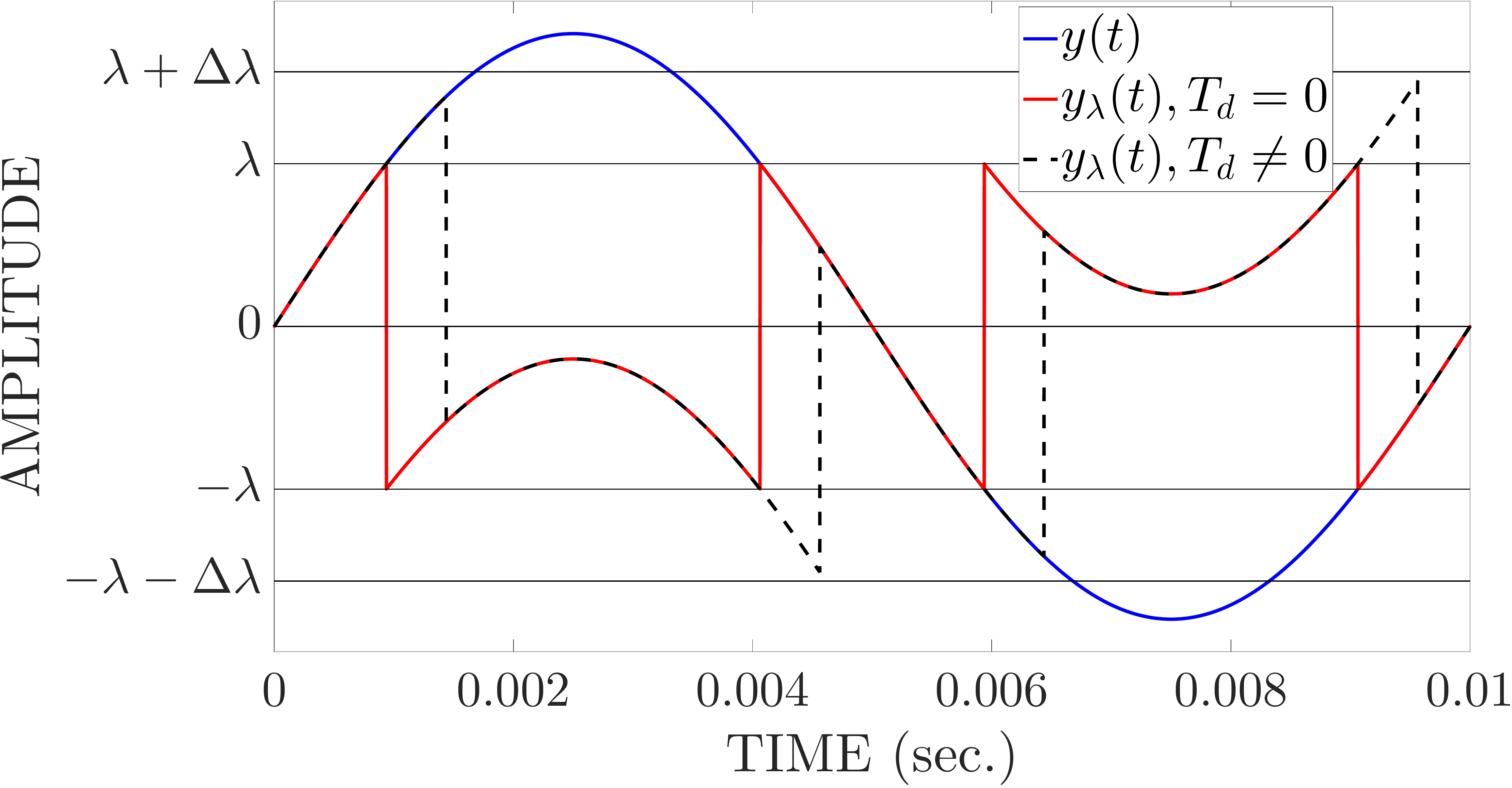}
	\caption{Effect of the time delay of the feedback loop on the modulo operation.}
	\label{Fig:mod_td}
\end{figure}

\subsection{Artifacts During Folding}
Errors or other artifacts that arise during folding operations in hardware result from various reasons. One of the major issues that arise in a modulo ADC is the time delay in the feedback loop (See Fig.~\ref{Fig:folding}). To elaborate, consider a scenario where $y(t)<\lambda$ for $t<t_1$ and it crosses $\lambda$ at time $t_1$. To fold the output voltage to the dynamic range of the ADC, $z(t) = 2\lambda u(t-t_1)$ needs to be subtracted from $y(t)$. However, there is a finite delay between the trigger time $t_1$ to generating $z(t)$. If the time delay is $T_d$, then $z(t) = 2\lambda u(t-t_1-T_d)$ is subtracted, which causes distortion. To illustrate this effect, in Fig.~\ref{Fig:mod_td}, we considered a sinusoidal signal (in blue) and its folded versions with and without delay. We observe that in the absence of any time delay ($T_d=0$) the signal folds perfectly (shown in red) to stay within the dynamic range. However, for a non-zero value of $T_d$, foldings do not take place at the folding instants, and the output of the modulo operator (shown in black) still remains outside the dynamic range $[-\lambda, \lambda]$, which results in a clipped modulo operator output. The proposed hardware solution addresses this issue and avoids the undesired clipping as shown in Figs.~\ref{Fig:Analysis_of_folding_ability} and \ref{Fig:frequency_response}.

Our solution uses the fact that with a finite time delay, the amount of overshoot of a smooth signal can be bounded. To elaborate, assume that the signal that undergoes the modulo operation is Lipschitz continuous. Specifically, a signal $y(t)$ is Lipschitz continuous if there exists a positive real number $\mathrm{L}_y$ such that for any $T>0$ we have
\begin{align}
    |y(t) -y(t+T)|\leq \mathrm{L}_y\, T.
    \label{eq:lipschitz}
\end{align}
With the Lipschitz smoothness condition, we note that the amplitude of $y(t)$ can not change more than $\mathrm{L}_y\, T_d$ between any folding instant and time of its effect to take place. Hence, if we choose the dynamic range of the ADC to be $[-(\lambda+\Delta \lambda), (\lambda+\Delta \lambda)]$ where $\Delta \lambda = \mathrm{L}_y\, T_d$ then the signal will not clip. We show this extended dynamic region in the example in Fig.~\ref{Fig:mod_td}. Alternatively, instead of increasing the ADC's dynamic range, one can keep it to be $[-\lambda, \lambda]$ and reduce the threshold values for comparators. In this case, Comp-1 will trigger when the input crosses $\lambda-\Delta \lambda$, and Comp-2 will trigger when the input goes beyond $-\lambda+\Delta \lambda$. In this way, the time delay issue is addressed by the modulo circuit without altering ADC's dynamic range. In our hardware, we choose the former solution. Specifically, we used an oscilloscope to measure and display samples. The dynamic range of the oscilloscope's ADC was sufficiently higher than $[-\lambda, \lambda]$ to sample signals of interest without clipping signals due to the delay in the feedback loop. 

In order to apply the solution, the signal must be Lipschitz continuous. In our design, the modulo operation input signal is always a bandlimited signal satisfying the Lipschitz smoothness condition \cite{bl_bound}. For a bandlimited signal $y(t)$, its Lipschitz constant $\mathrm{L}_y$ is directly proportional to its bandwidth $\omega_c$ \cite{bl_bound}. Hence, for a given value $\Delta \lambda$ and $T_d$ (both depend on the modulo circuit), $\mathrm{L}_y = \Delta \lambda/T_d$ is fixed and this restricts the maximum frequency of the input signal that can be faithfully folded. In the current design, we choose to implement the counter management using a TEENSY microcontroller, resulting in a $1 \mu$s time delay. Then for a sinusoidal signal $y(t) = A \sin(2\pi f_0 t)$, the Lipschitz constant is given as $L_y = 2\pi A f_0$. For $A = 8\lambda$, $\Delta \lambda = 0.5 \lambda$, and $T_d = 1 \mu$s, we note that the maximum operating frequency is 10 kHz which is in line with the experimental results discussed in Fig. \ref{Fig:frequency_response}.



%


\section{Results}
In this section, we demonstrate the modulo hardware's signal reconstruction capability. We focus on the folding and reconstruction of bandlimited and FRI signals. In the hardware, the folded measurements are generally  contaminated by different noises, including quantization noise. Since the performance of an unfolding algorithm depends on the noise levels, we first discuss a few simulated results to assess the performance of the $B^2R^2$ algorithm used for unfolding. Then we demonstrate the results from the hardware.

\begin{figure}[!t]
	\centering
	\includegraphics[width= 2.5 in]{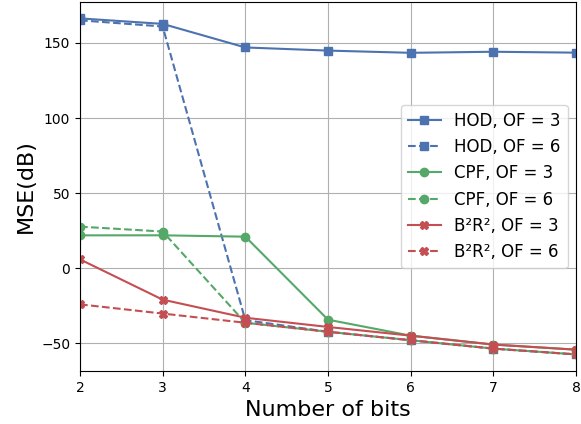}
	\caption{Comparison of HOD, CPF, and $B^2R^2$ algorithms after the quantization process in terms of MSE when recovering a bandlimited signal from modulo samples with $\lambda = 1.25$ and OF = 3,6. For a given number of bits, $B^2R^2$ has the lowest MSE.}
	\label{Fig:quantization}
\end{figure}

\begin{figure}[!t]
	\centering
	\includegraphics[width= 2.5 in]{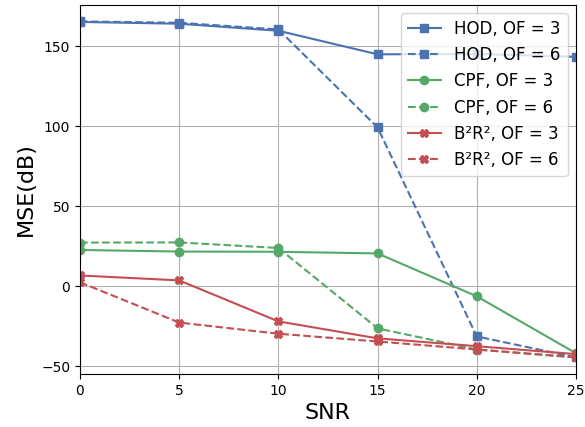}
	\caption{Comparison of HOD, CPF and $B^2R^2$ algorithms (with unbounded noise) in terms of MSE when recovering a bandlimited signal from modulo samples with  $\lambda$ = 1.25 and OF = 3,6. For a given SNR and OF, $B^2R^2$ has the lowest MSE.}
	\label{Fig:unbounded}
\end{figure}
\subsection{Simulated Results}
In this section, we compare our $B^2R^2$ algorithm with the \emph{higher-order differences} (HOD) approach \cite{unlimited_sampling17, uls_tsp} and \emph{Chebyshev polynomial filter}-based (CPF) method \cite{uls_romonov}. Although a comparison of these methods is analyzed in \cite{eyar_moduloicassp}, the settings are different here. Importantly, quantization noise is not considered in our previous work.

We consider the noisy measurements as 
\begin{align}
            \tilde{y}_{\lambda}(n T_s) = {y}_{\lambda}(n T_s) + v(n T_s),
        \label{eq: noisy_samples}
        \end{align}
where $v(n T_s)$ is the noise term. In the simulations, $\lambda$ is set to be $1.25$ as in the hardware. We normalize the bandlimited signals to have a maximum amplitude of $10$. The SNR is calculated as $\text{SNR} = 20 \log\left(\frac{||y_{\lambda}(n T_s)||}{||v(n T_s)||}\right)$. Reconstruction accuracies of different methods are compared in terms of the normalized mean-squared errors (MSEs) as $ \frac{\sum |y(nT_s) - \hat{y}(n Ts)|^2}{\sum |y(nT_s)|^2}$, where $\hat{y}(n T_s)$ is an estimate of $y(n T_s)$. For different noise settings and over-sampling factors (OFs), we considered 100 independent noise realizations and calculated the average MSE for them. We first consider quantization noise and then present results for unbounded noise.
\begin{figure}[!t]
    \centering
    \includegraphics[width=0.3\textwidth]{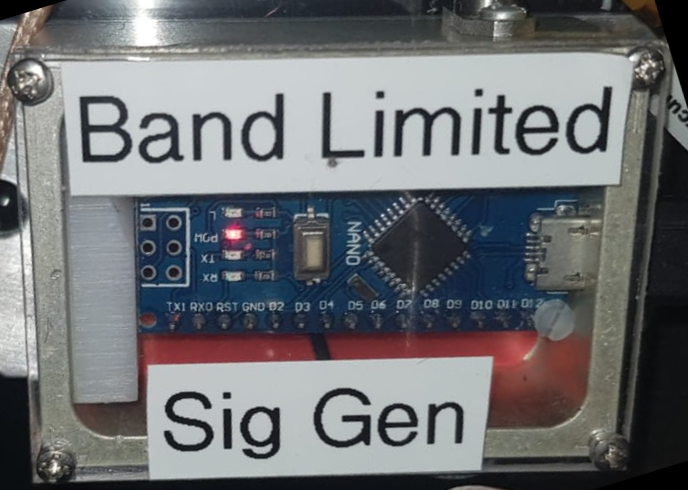}
    \caption{Bandlimited signal generator.}
    \label{fig:UDRSignalGen}
\end{figure}


 In the first simulation, the unfolding algorithms are applied to quantized folded samples. The MSE in the estimation of bandlimited signals for a different number of bits and OFs is shown in Fig. ~\ref{Fig:quantization}. We observe that for a given $\text{OF}$, $B^2R^2$ algorithm results in the lowest MSE for less than 5 bits. For more than five bits, all the algorithms, except HOD with OF = 3, perform equally well. The results show that low-resolution quantizers can be used with the $B^2R^2$ algorithm for unfolding, which saves both power and memory requirements.

Next, for unbounded noise, we assume that the noise samples $v(nT_s)$ are independent and identically distributed Gaussian random variables with zero means. The variance of $v(nT_s)$ is set to achieve the desired SNR. We compare the methods for different values of SNR and OFs with $\lambda = 1.25$. Fig.~\ref{Fig:unbounded} shows the MSE of the different algorithms for OF $= 3$ and $6$. We note that our algorithm results in the lowest error for a given OF and SNR.

Given the advantages of the $B^2R^2$ algorithm over the other approaches, we present the hardware results in the next section by using this method.

\begin{figure}[!t]
	\centering
	\begin{tabular}{c}
		\subfigure[]{\includegraphics[width=3.2in]{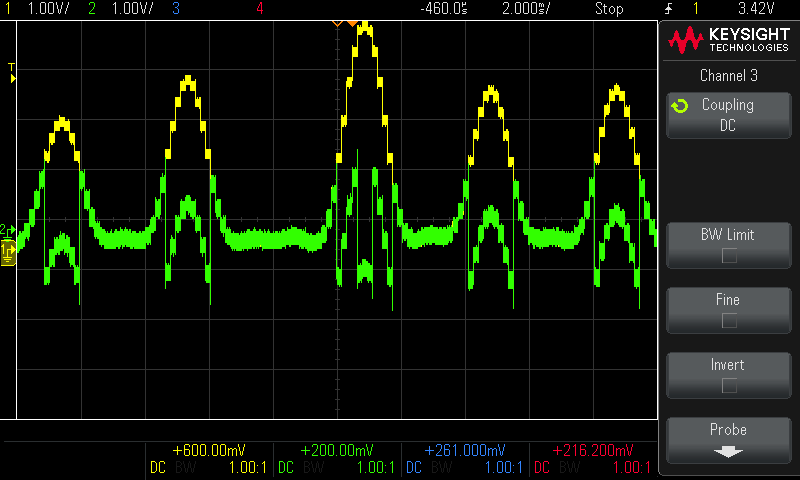}\label{fig:BL1_scope}} \\
		\subfigure[]{\includegraphics[width=3.2in]{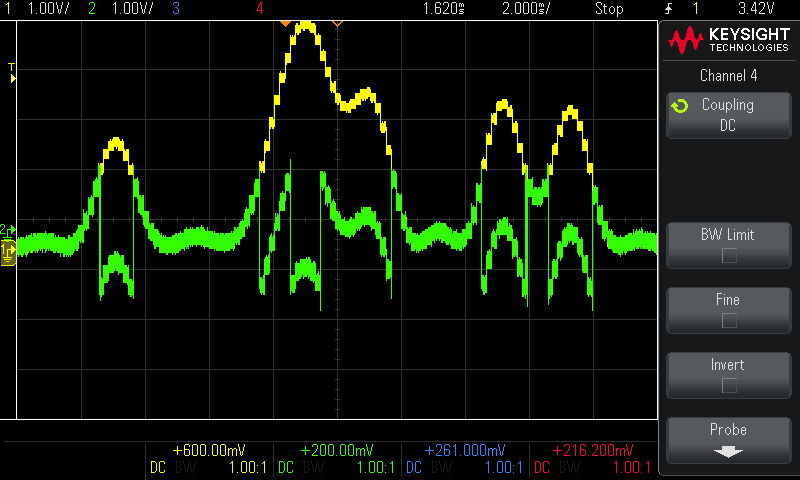}\label{fig:BL2_scope}} 
	\end{tabular}
	\caption{Screenshots of the oscilloscope capturing the bandlimited input signals (yellow) and its folded output signals (green).}
	\label{fig:BL_scopeex1}
\end{figure}
\begin{figure}[!t]
	\centering
	\begin{tabular}{c}
		\subfigure[]{\includegraphics[width=3.2in]{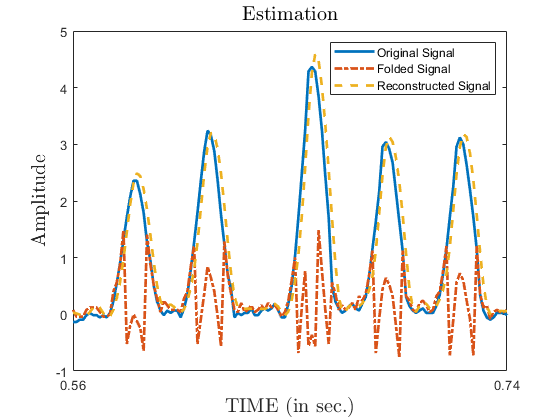}\label{fig:BL1}} \\
		\subfigure[]{\includegraphics[width=3.2in]{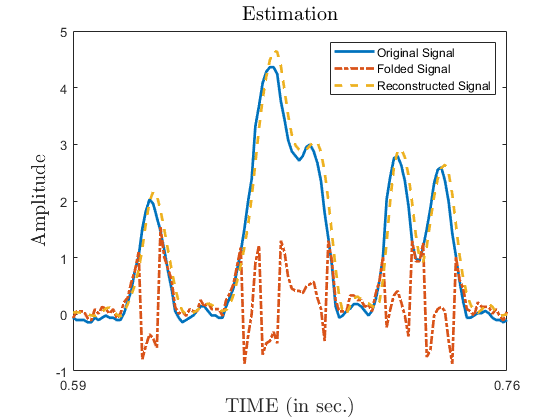}\label{fig:BL2}} 
	\end{tabular}
	\caption{Hardware results for bandlimited signals. The $B^2R^2$ algorithm is used to unfold $y_\lambda(t)$ (measured at the output of hardware), and the unfolded signal $\hat{y}(t)$ is plotted with bandlimited signal $y(t)$.}
	\label{fig:BL_ex1}
\end{figure}

\begin{figure}[!t]
	\centering
	\begin{tabular}{c}
		\subfigure[Screenshot of the oscilloscope capturing filtered FRI signal (yellow), its folded output (green), and the DVG signal (blue).] {\includegraphics[width=3.2in]{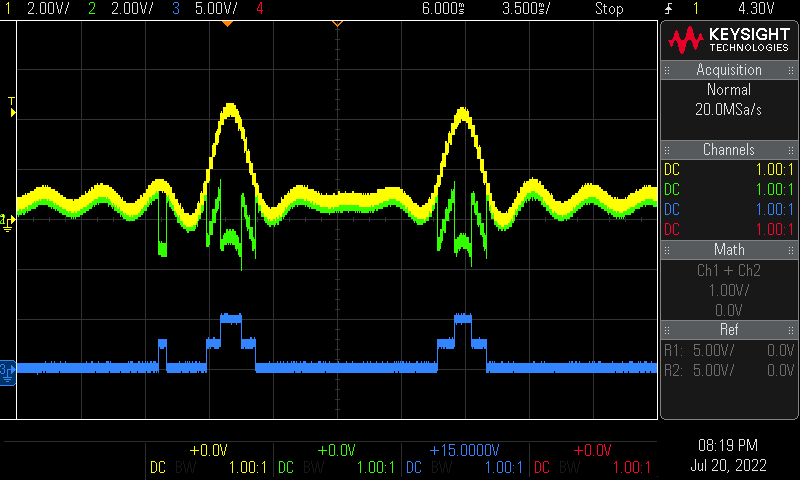}\label{fig:FRI1_1k}} \\
		\subfigure[Filtered FRI signal with its folded and unfolded versions.]{\includegraphics[width=3.2in]{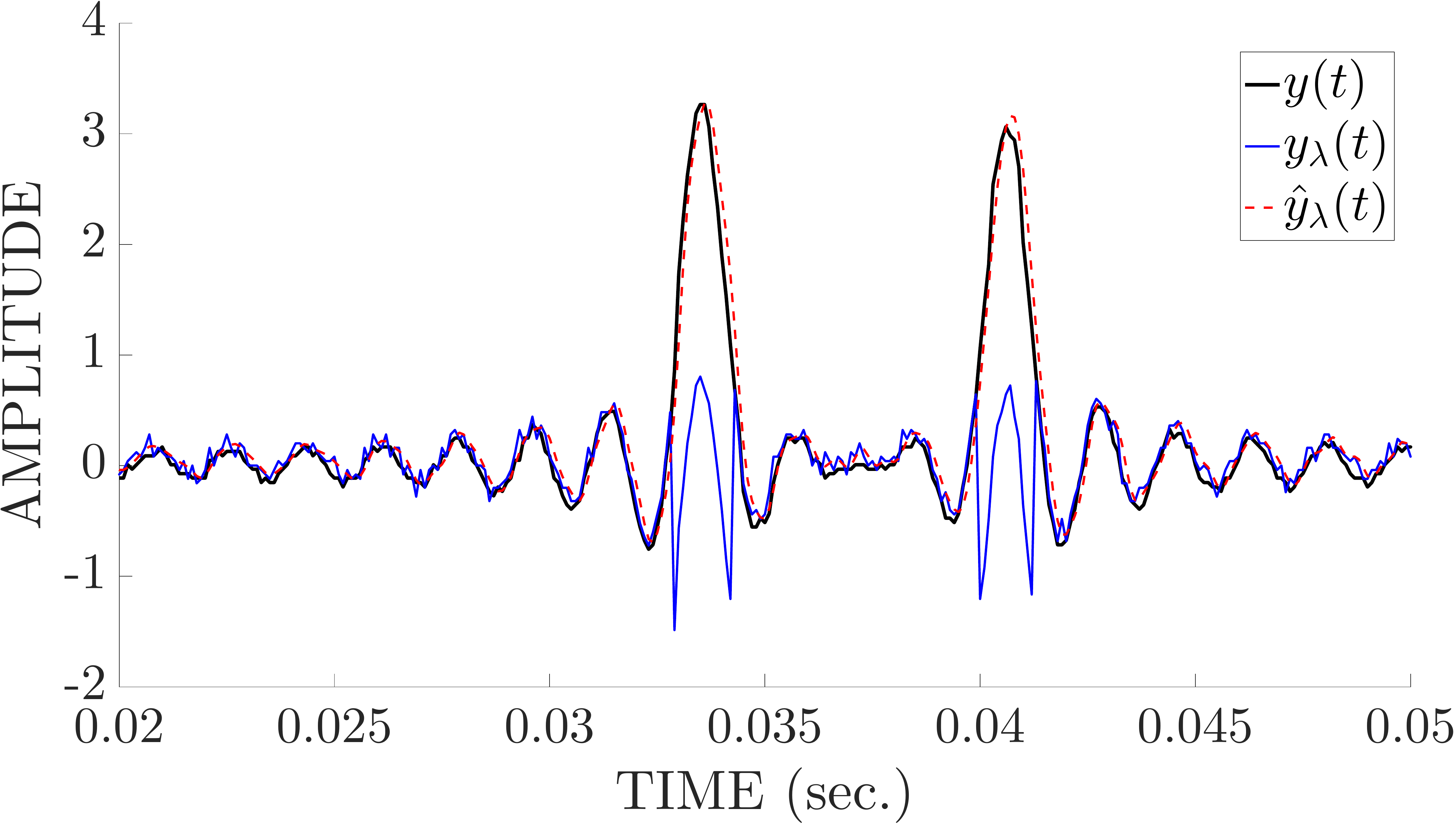}\label{fig:FRI_LPF1}}\\
		\subfigure[FRI signal reconstruction.] {\includegraphics[width=3.2in]{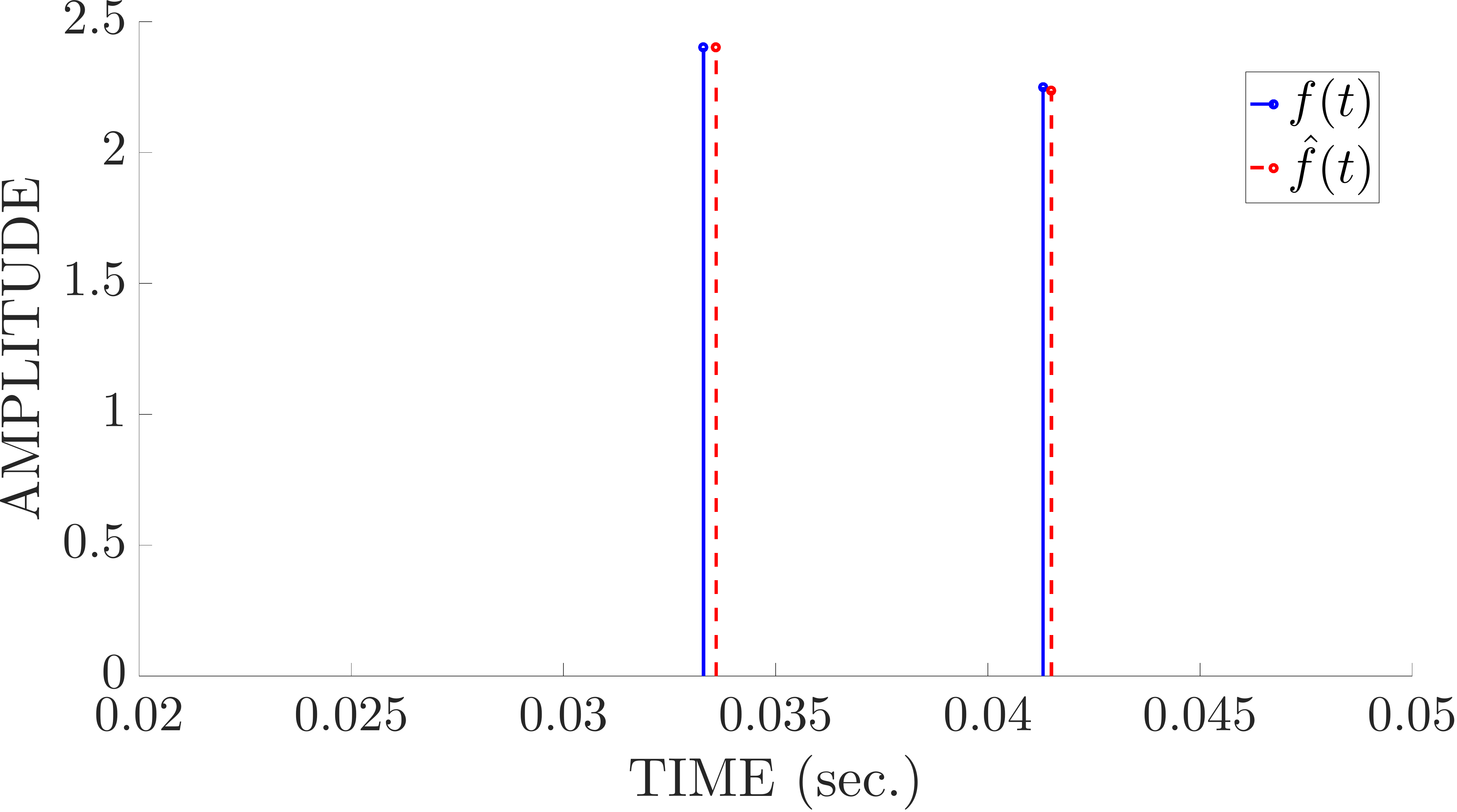}\label{fig:FRI1_REC}} \\
		
	\end{tabular}
	\caption{Reconstruction of FRI signal ($L=2$) via the modulo hardware.}
	\label{Fig:FRI1}
\end{figure}

\begin{figure}[!t]
	\centering
	\begin{tabular}{c}
		\subfigure[Screenshot of the oscilloscope capturing filtered FRI signal (yellow), its folded output (green), and the DVG signal (blue).] {\includegraphics[width=3.2in]{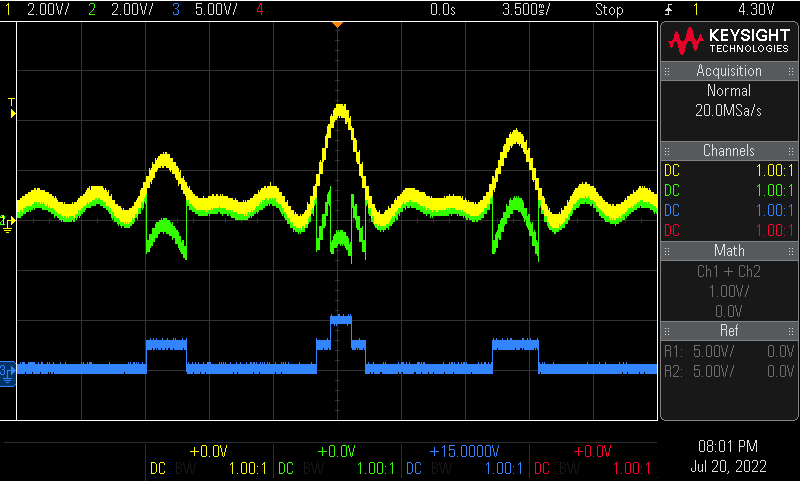}\label{fig:FRI2_1k}} \\
		\subfigure[Filtered FRI signal with its folded and unfolded versions.]{\includegraphics[width=3.2in]{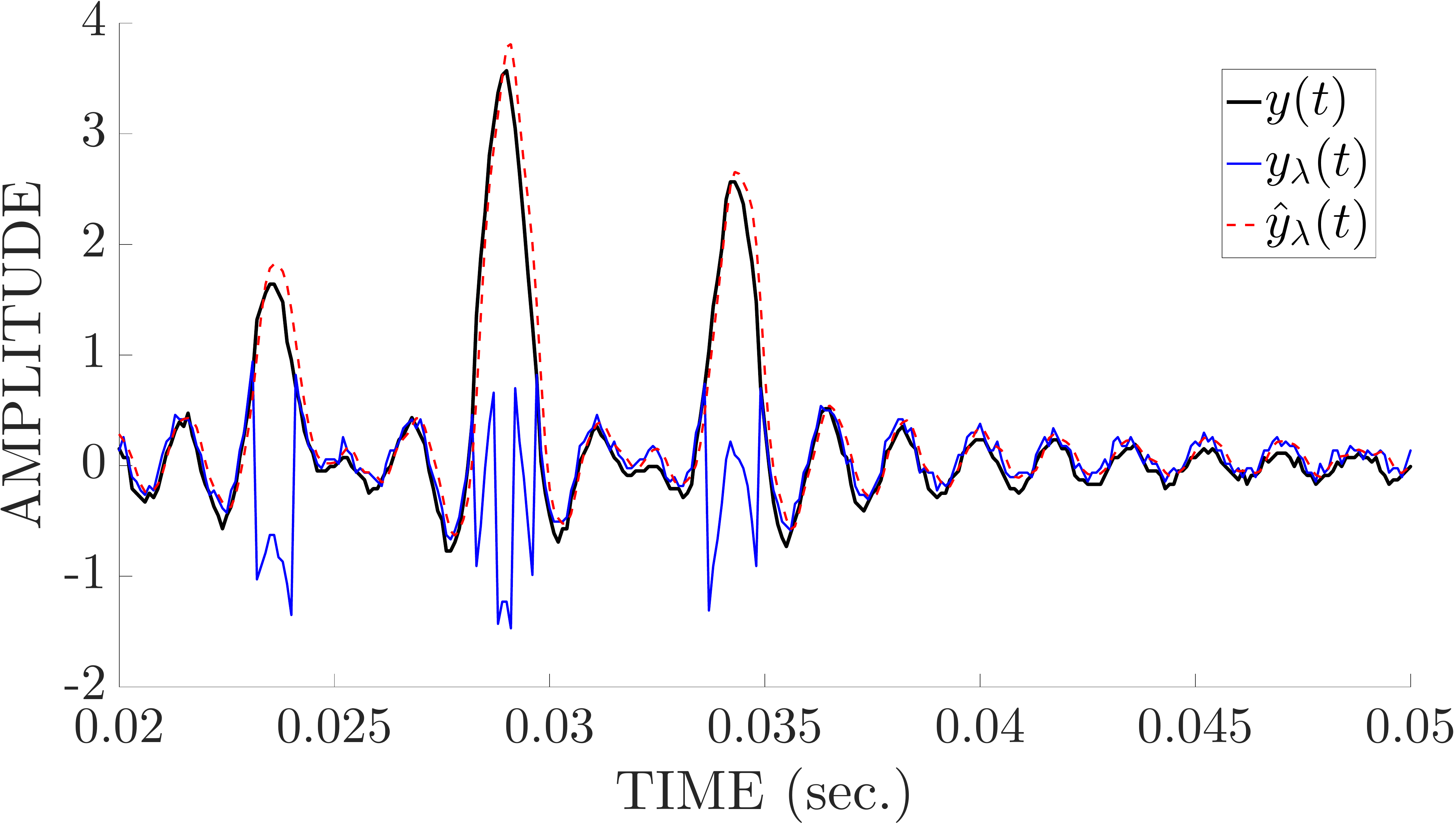}\label{fig:FRI_LPF2}}\\
		\subfigure[FRI signal reconstruction.] {\includegraphics[width=3.2in]{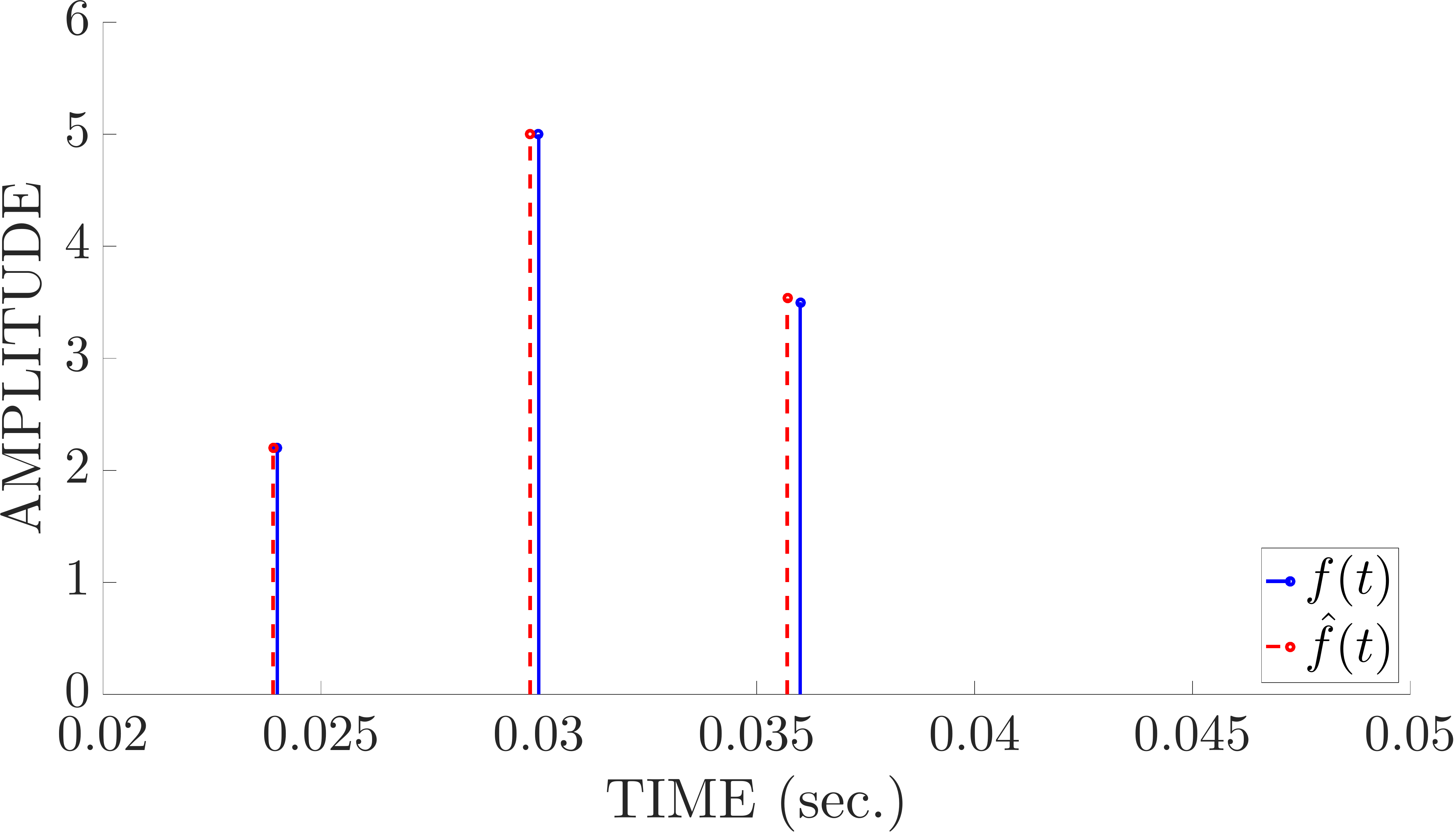}\label{fig:FRI2_REC}} \\
	\end{tabular}
	\caption{Reconstruction of FRI signal ($L=3$) via the modulo hardware.}
	\label{Fig:FRI2}
\end{figure}

\begin{figure}[!t]
	\centering
	\begin{tabular}{c}
		\subfigure[Screenshot of the oscilloscope capturing filtered FRI signal (yellow), its folded output (green), and the DVG signal (blue).] {\includegraphics[width=3.2in]{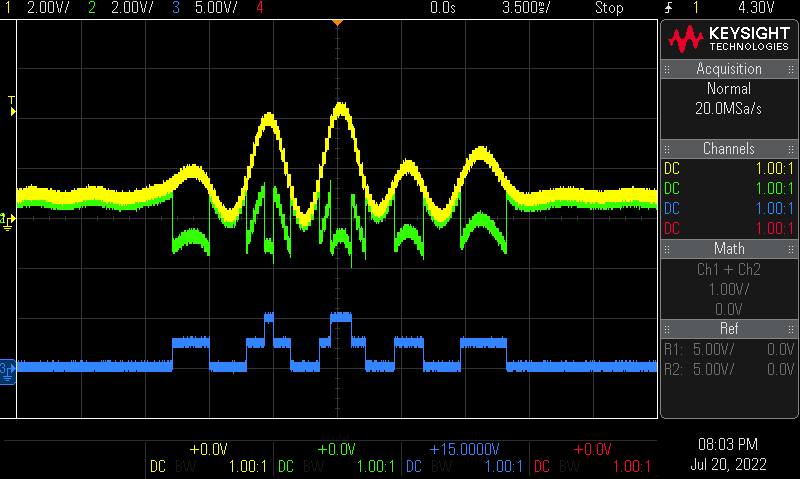}\label{fig:FRI3_1k}} \\
		\subfigure[Filtered FRI signal with its folded and unfolded versions.]{\includegraphics[width=3.2in]{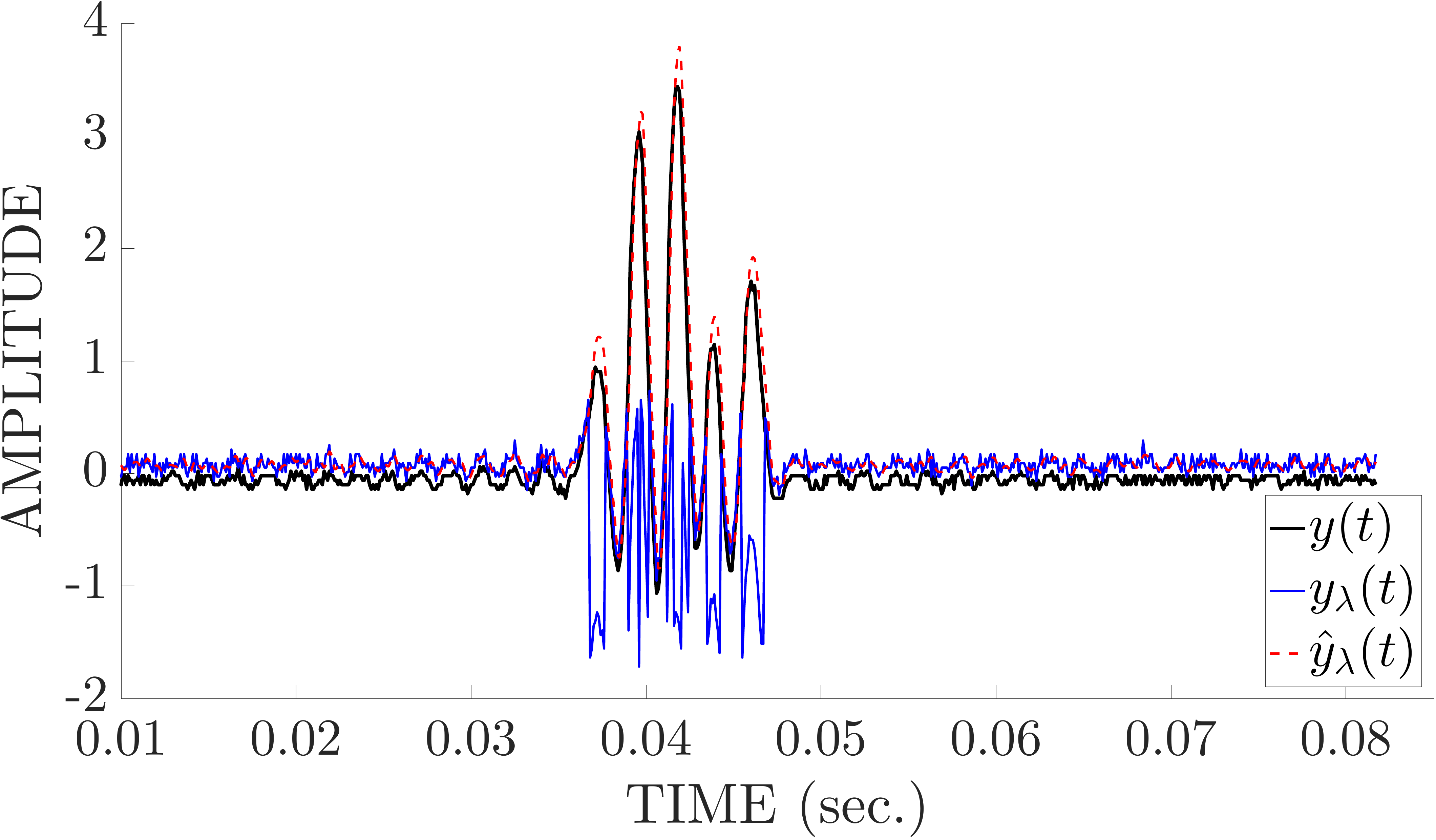}\label{fig:FRI_LPF3}}\\
		\subfigure[FRI signal reconstruction.] {\includegraphics[width=3.2in]{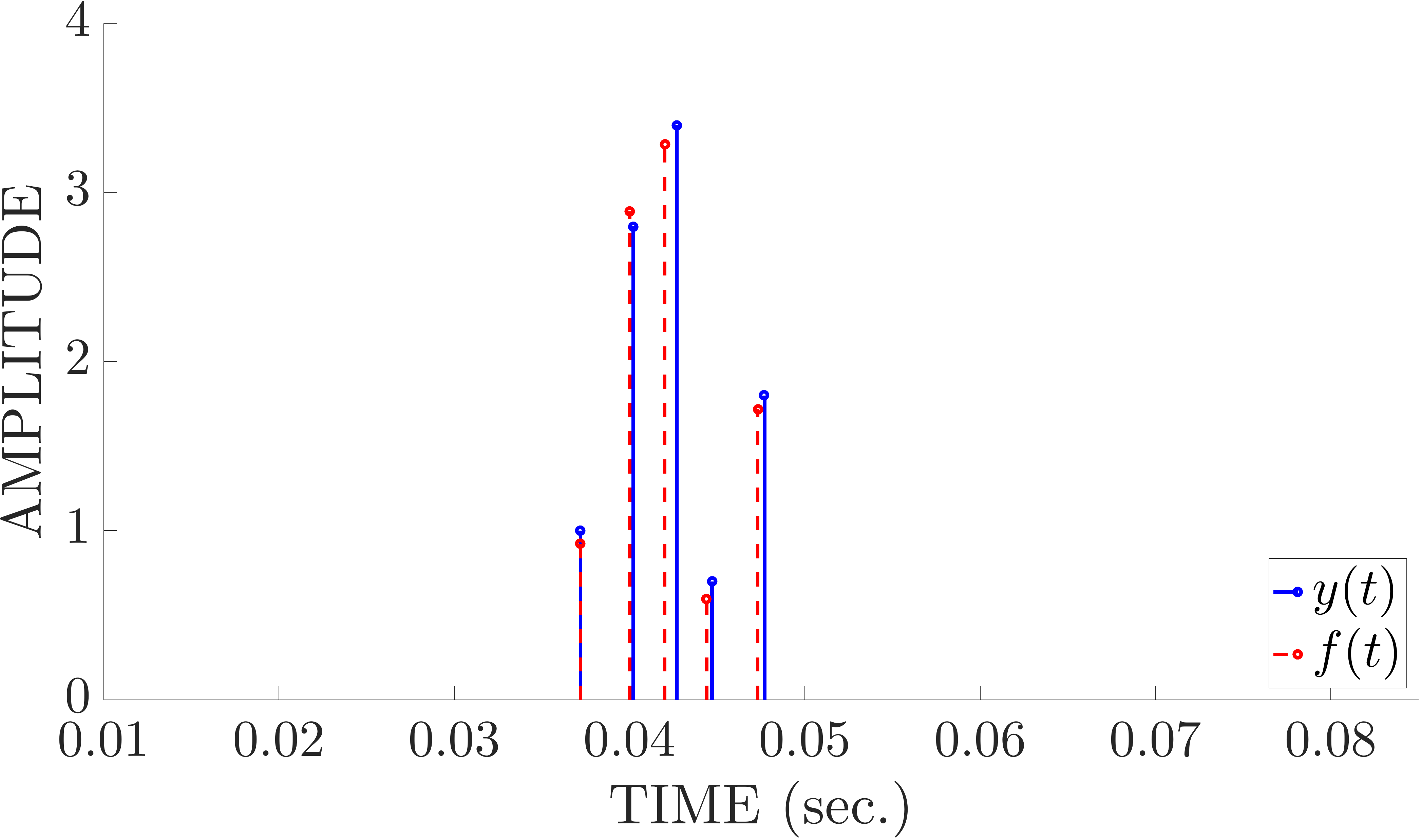}\label{fig:FRI3_REC}} \\
		
	\end{tabular}
	\caption{Reconstruction of FRI signal ($L=5$) via the modulo hardware.}
	\label{Fig:FRI3}
\end{figure}


\subsection{Hardware Results for Bandlimited Signals}
In this section, we first present results for bandlimited signals. For generating bandlimited or lowpass signals, we used an Arduino microcontroller (See Fig.~\ref{fig:UDRSignalGen}) which converts the digital signal to an analog signal via a DAC. The digital signals were generated using MATLAB software. Two examples of 1khz bandlimited signal are presented in Fig. \ref{fig:BL1_scope} and Fig. \ref{fig:BL2_scope}. The modulo hardware folds the signals to stay within the dynamic range, as shown in Fig.~\ref{fig:BL_scopeex1}. The signals are sampled with an oversampling factor of five ($\text{OF}=5$), and the $B^2R^2$ algorithm is applied for unfolding. The unfolded or reconstructed signals are shown in Fig.~\ref{fig:BL_ex1}. We observe that the reconstruction is close to the true signals except for an amplitude scaling factor, which is the result of scaling within the hardware.

\subsection{Hardware Results for FRI Signals}
Before presenting the results for FRI signals, we briefly discuss the FRI signal model and its sampling and reconstruction mechanism for ease of discussion. 
Consider an FRI signal consisting of a stream of $L$ pulses:  
\begin{align}
	f(t) = \sum_{\ell=1}^L a_\ell h(t-t_\ell),
	\label{eq:fri1}
\end{align}
where the pulse $h(t)$ a real-valued known pulse. We assume that $\{a_\ell\}_{\ell = 1}^L$ are real-valued and $\{t_\ell\}_{\ell = 1}^L \subset (0, T_0] \subset \mathbb{R}$ for a known $T_0$.

The FRI signal model in \eqref{eq:fri1} is encountered in several scientific applications such as radar imaging \cite{bar_radar, bajwa_radar, sunil}, ultrasound imaging \cite{eldar_sos,eldar_beamforming,mulleti_icip}, light detection and ranging (LIDAR) \cite{castorena_lidar}, time-domain optical coherence tomography (TDOCT) \cite{blu_oct}, and other time-of-flight imaging systems. In these applications,  $h(t)$ is the transmitted pulse and $\{a_\ell \,h(t-t_\ell)\}_{\ell=1}^{L}$ constitute the reflections from $L$ point targets. The amplitudes $\{a_\ell\}_{\ell = 1}^L$ depend on the sizes of the targets and the delays $\{t_\ell\}_{\ell = 1}^L$ are proportional to the distances of the targets from the transmitter. Here $T_0$ denotes the maximum time delay of the targets. The signal $f(t)$ is specified by $\{a_\ell, t_\ell\}_{\ell=1}^L$ and can be reconstructed from its sub-Nyquist measurements acquired using an appropriate sampling kernel \cite{vetterli,eldar_sos,fri_strang, mulleti_kernal}. Given their widespread application, here we consider sampling and reconstruction of FRI signals by using our hardware prototype.

FRI signals can be perfectly reconstructed by applying high-resolution spectral estimation methods, such as the annihilating filter (AF) or Prony's method and its variants
\cite{prony, plonka, schmidt_music, barabell_rmusic, bdrao_rmusic,sarkar_mp,paulraj_esprit} to the Fourier measurements
\begin{align}
	S(k\omega_0) = \frac{F(k\omega_0)}{H(k\omega_0)} =  \sum_{\ell = 1}^L a_\ell \, e^{-\mathrm{j}k\omega_0 t_\ell}, \, k \in \{-K, \cdots, K\},
	\label{eq:Skomega}
\end{align}
where we assume that $H(k \omega_0) \neq 0$. Here $K \geq L$ and $\omega_0 = \frac{2\pi}{T_0}$ \cite{eldar_2015sampling}. The Fourier measurements $\{S(k\omega_0)\}_{k =-K}^K$ can be determined from the samples $(f*g)(nT_s)$ where $g(t)$ is an ideal lowpass filter with bandwidth $[-K\omega_0, K \omega_0]$ and $T_s = \frac{2\pi}{(2K+1)\omega_0}$. In practice, the duration of the pulse $h(t)$ is very short, and hence $f(t)$ has a wide bandwidth. This results in a large sampling rate (or Nyquist rate) if $f(t)$ is sampled directly. However, the filtered signal $y(t) = (f*g)(t)$ is bandlimited to $[-K\omega_0, K \omega_0]$, which is much smaller than that of $h(t)$ and the sampling rate is much lower than the Nyquist rate. 

As in the bandlimited signal model, a modulo operation can be applied to the filtered signal $y(t)$ to avoid clipping. Then $y_\lambda(t)$ is sampled. Then to determine the Fourier samples $\{S(k\omega_0)\}_{k =-K}^K$, unfolding is first applied. Since the filtered signal is bandlimited, we use the proposed hardware for modulo folding and can apply the $B^2R^2$ algorithm from unfolding.

In our setup, to generate the FRI signals, we consider $h(t)$ to be a short pulse of bandwidth 30kHz (Nyquist rate = 60kHz). We consider three examples with $L=2, 3,$ and 5. The amplitudes and time delays are generated randomly. The maximum time delay is $T_0 = 0.1$ sec. Once generated, the FRI signal is lowpass filtered with a cutoff frequency of 1kHz. MATLAB is used to generate the samples of filtered FRI signals and then an Arduino microcontroller is used to generate the analog counterpart of them. The signal is then folded using the hardware, and the folded signals are sampled. The sampling rate is 10kHz which is five times higher than the sampling rate of the lowpass signal. Still, the rate is six times lower than the Nyquist rate, and hence the system operates at a sub-Nyquist rate. 


We first applied the $B^2R^2$ algorithm to unfold the signal and then used ESPRIT \cite{paulraj_esprit} to estimate the time delays and amplitudes of the FRI signals. 
In Fig.~\ref{Fig:FRI1}, Fig.~\ref{Fig:FRI2} and  Fig.~\ref{Fig:FRI3} we show sampling and reconstruction of FRI signals with $L=2,3,5$ respectively.
The FRI signals followed by a lowpass filter are presented in Fig.~\ref{fig:FRI1_1k}, Fig.~\ref{fig:FRI2_1k}, and  Fig.~\ref{fig:FRI3_1k}.
The reconstruction of the lowpass signals displayed in Fig.~\ref{fig:FRI_LPF1}, Fig.~\ref{fig:FRI_LPF2} and  Fig.~\ref{fig:FRI_LPF3}, where $y(t)$ described the LPF output, $y_\lambda(t)$ is the folded signal (output of the modulo hardware), and the unfolded signals are given by $\hat{y}(t)$. Fig.~\ref{fig:FRI1_REC}, Fig.~\ref{fig:FRI2_REC} and Fig.~\ref{fig:FRI3_REC} show location and amplitude of the true signal $f(t)$ and estimated FRI signal $\hat{f}(t)$. 
The maximum error in the estimation of time delay is $-15$ dB which shows that the system can be used in applications like radar and ultrasound imaging.

\section{Conclusions}
 We presented a hardware prototype for the modulo folding system and showed that for different bandlimited and FRI signals, the hardware is able to fold the signal faithfully. In particular, we were able to sample signals with 8 times the dynamic range of the ADC roughly. We also  addressed the time delay issue of the modulo system and presented a hardware solution. The overall system operates five times below the Nyquist rate, which enables one to use low-rate, low-dynamic range, power-efficient ADCs.

\bibliographystyle{IEEEtran}
\bibliography{refs,refs2,US_biblios}

\begin{thebibliography}{10}
\providecommand{\url}[1]{#1}
\csname url@samestyle\endcsname
\providecommand{\newblock}{\relax}
\providecommand{\bibinfo}[2]{#2}
\providecommand{\BIBentrySTDinterwordspacing}{\spaceskip=0pt\relax}
\providecommand{\BIBentryALTinterwordstretchfactor}{4}
\providecommand{\BIBentryALTinterwordspacing}{\spaceskip=\fontdimen2\font plus
\BIBentryALTinterwordstretchfactor\fontdimen3\font minus
  \fontdimen4\font\relax}
\providecommand{\BIBforeignlanguage}[2]{{%
\expandafter\ifx\csname l@#1\endcsname\relax
\typeout{** WARNING: IEEEtran.bst: No hyphenation pattern has been}%
\typeout{** loaded for the language `#1'. Using the pattern for}%
\typeout{** the default language instead.}%
\else
\language=\csname l@#1\endcsname
\fi
#2}}
\providecommand{\BIBdecl}{\relax}
\BIBdecl

\bibitem{marks_clipping2}
R.~Marks, ``Restoring lost samples from an oversampled band-limited signal,''
  \emph{IEEE Trans. Acoust., Speech, Signal Process.}, vol.~31, no.~3, pp.
  752--755, 1983.

\bibitem{marks_clipping1}
R.~Marks and D.~Radbel, ``Error of linear estimation of lost samples in an
  oversampled band-limited signal,'' \emph{IEEE Trans. Acoust., Speech, Signal
  Process.}, vol.~32, no.~3, pp. 648--654, 1984.

\bibitem{perez2011automatic}
J.~P.~A. P{\'e}rez, S.~C. Pueyo, and B.~C. L{\'o}pez, \emph{Automatic gain
  control}.\hskip 1em plus 0.5em minus 0.4em\relax Springer, 2011.

\bibitem{mercy1981review}
D.~Mercy, ``A review of automatic gain control theory,'' \emph{Radio and
  Electronic Engineer}, vol.~51, no. 11.12, pp. 579--590, 1981.

\bibitem{landau_compander}
H.~J. Landau, ``On the recovery of a band-limited signal, after instantaneous
  companding and subsequent band limiting,'' \emph{The Bell System Technical
  Journal}, vol.~39, no.~2, pp. 351--364, 1960.

\bibitem{landau_distorted_bl}
H.~J. Landau and W.~L. Miranker, ``The recovery of distorted band-limited
  signals,'' \emph{J. Mathematical Anal. Appl.}, vol.~2, no.~1, pp. 97--104,
  1961.

\bibitem{uls_tsp}
A.~{Bhandari}, F.~{Krahmer}, and R.~{Raskar}, ``On unlimited sampling and
  reconstruction,'' \emph{IEEE Trans. Signal Process.}, vol.~69, pp.
  3827--3839, 2020.

\bibitem{uls_romonov}
E.~{Romanov} and O.~{Ordentlich}, ``Above the {N}yquist rate, modulo folding
  does not hurt,'' \emph{IEEE Signal Process. Lett.}, vol.~26, no.~8, pp.
  1167--1171, 2019.

\bibitem{bhandari2021unlimited}
A.~Bhandari, F.~Krahmer, and T.~Poskitt, ``Unlimited sampling from theory to
  practice: {F}ourier-{P}rony recovery and prototype {ADC},'' \emph{IEEE Trans.
  Signal Process.}, vol.~70, pp. 1131--1141, 2022.

\bibitem{eyar_moduloicassp}
E.~Azar, S.~Mulleti, and Y.~C. Eldar, ``Residual recovery algorithm for modulo
  sampling,'' in \emph{Proc. Intl. Conf. Acoust., Speech and Signal Process.
  (ICASSP)}, 2022, pp. 5722--5726.

\bibitem{azar2022robust}
------, ``Robust unlimited sampling beyond modulo,'' \emph{arXiv preprint
  arXiv:2206.14656}, 2022.

\bibitem{uls_fri}
A.~Bhandari, F.~Krahmer, and R.~Raskar, ``Unlimited sampling of sparse
  signals,'' in \emph{Proc. Intl. Conf. Acoust., Speech and Signal Process.
  (ICASSP)}, 2018, pp. 4569--4573.

\bibitem{uls_sparsevec}
D.~Prasanna, C.~Sriram, and C.~R. Murthy, ``On the identifiability of sparse
  vectors from modulo compressed sensing measurements,'' \emph{IEEE Signal
  Process. Lett.}, vol.~28, pp. 131--134, 2021.

\bibitem{uls_doa}
S.~Fern\'andez-Menduiña, F.~Krahmer, G.~Leus, and A.~Bhandari, ``{DoA}
  estimation via unlimited sensing,'' in \emph{Proc. European Signal Process.
  Conf. (EUSIPCO)}, 2021, pp. 1866--1870.

\bibitem{uls_radon}
A.~Bhandari, M.~Beckmann, and F.~Krahmer, ``The modulo {R}adon transform and
  its inversion,'' in \emph{Proc. European Signal Process. Conf. (EUSIPCO)},
  2021, pp. 770--774.

\bibitem{uls_graph}
F.~Ji, P.~Pratibha, and W.~P. Tay, ``On folded graph signals,'' in \emph{Proc.
  Global Conf. Signal Info. Process. (GlobalSIP)}, 2019, pp. 1--5.

\bibitem{sradc_park}
D.~Park, J.~Rhee, and Y.~Joo, ``A wide dynamic-range {CMOS} image sensor using
  self-reset technique,'' \emph{IEEE Electron Device Lett.}, vol.~28, no.~10,
  pp. 890--892, 2007.

\bibitem{sradc_sasagwa}
K.~Sasagawa, T.~Yamaguchi, M.~Haruta, Y.~Sunaga, H.~Takehara, H.~Takehara,
  T.~Noda, T.~Tokuda, and J.~Ohta, ``An implantable {CMOS} image sensor with
  self-reset pixels for functional brain imaging,'' \emph{IEEE Trans. Electron
  Devices}, vol.~63, no.~1, pp. 215--222, 2016.

\bibitem{sradc_yuan}
J.~Yuan, H.~Y. Chan, S.~W. Fung, and B.~Liu, ``An activity-triggered 95.3 db
  {DR} $-$75.6 db {THD CMOS} imaging sensor with digital calibration,''
  \emph{IEEE J. Solid-State Circuits}, vol.~44, no.~10, pp. 2834--2843, 2009.

\bibitem{krishna2019unlimited}
A.~Krishna, S.~Rudresh, V.~Shaw, H.~R. Sabbella, C.~S. Seelamantula, and C.~S.
  Thakur, ``Unlimited dynamic range analog-to-digital conversion,'' \emph{arXiv
  preprint:1911.09371}, 2019.

\bibitem{Bhandari:2022:J}
A.~Bhandari, ``Back in the {US}-{SR}: {Unlimited} sampling and sparse
  super-resolution with its hardware validation,'' vol.~29, pp. 1047--1051,
  Mar. 2022.

\bibitem{mod_hyst}
D.~Florescu, F.~Krahmer, and A.~Bhandari, ``The surprising benefits of
  hysteresis in unlimited sampling: Theory, algorithms and experiments,''
  \emph{IEEE Trans. Signal Process.}, vol.~70, pp. 616--630, 2022.

\bibitem{vetterli}
M.~Vetterli, P.~Marziliano, and T.~Blu, ``Sampling signals with finite rate of
  innovation,'' \emph{IEEE Trans. Signal Process.}, vol.~50, no.~6, pp.
  1417--1428, Jun. 2002.

\bibitem{eldar_sos}
R.~Tur, Y.~C. Eldar, and Z.~Friedman, ``Innovation rate sampling of pulse
  streams with application to ultrasound imaging,'' \emph{IEEE Trans. Signal
  Process.}, vol.~59, no.~4, pp. 1827--1842, Apr. 2011.

\bibitem{mulleti_kernal}
S.~{Mulleti} and C.~S. {Seelamantula}, ``Paley--{W}iener characterization of
  kernels for finite-rate-of-innovation sampling,'' \emph{IEEE Trans. Signal
  Process.}, vol.~65, no.~22, pp. 5860--5872, Nov. 2017.

\bibitem{mulleti2022modulo}
S.~Mulleti and Y.~C. Eldar, ``Modulo sampling of {FRI} signals,'' \emph{arXiv
  preprint arXiv:2207.08774}, 2022.

\bibitem{bl_bound}
A.~Papoulis, ``Limits on bandlimited signals,'' \emph{Proc. IEEE}, vol.~55,
  no.~10, pp. 1677--1686, 1967.

\bibitem{unlimited_sampling17}
A.~Bhandari, F.~Krahmer, and R.~Raskar, ``On unlimited sampling,'' in
  \emph{Proc. Intl. Conf. Sampling theory and Appl. (SampTA)}, July 2017, pp.
  31--35.

\bibitem{bar_radar}
O.~Bar-Ilan and Y.~C. Eldar, ``Sub-{N}yquist radar via {D}oppler focusing,''
  \emph{IEEE Trans. Signal Process.}, vol.~62, no.~7, pp. 1796--1811, Apr.
  2014.

\bibitem{bajwa_radar}
W.~U. Bajwa, K.~Gedalyahu, and Y.~C. Eldar, ``Identification of parametric
  underspread linear systems and super-resolution radar,'' \emph{IEEE Trans.
  Signal Process.}, vol.~59, no.~6, pp. 2548--2561, Jun. 2011.

\bibitem{sunil}
S.~Rudresh and C.~S. Seelamantula, ``Finite-rate-of-innovation-sampling-based
  super-resolution radar imaging,'' \emph{IEEE Trans Signal Process.}, vol.~65,
  no.~19, pp. 5021--5033, 2017.

\bibitem{eldar_beamforming}
N.~Wagner, Y.~C. Eldar, and Z.~Friedman, ``Compressed beamforming in ultrasound
  imaging,'' \emph{IEEE Trans. Signal Process.}, vol.~60, no.~9, pp.
  4643--4657, Sep. 2012.

\bibitem{mulleti_icip}
S.~Mulleti, S.~Nagesh, R.~Langoju, A.~Patil, and C.~S. Seelamantula,
  ``Ultrasound image reconstruction using the finite-rate-of-innovation
  principle,'' in \emph{Proc. IEEE Int. Conf. Image Process. (ICIP)}, Oct.
  2014, pp. 1728--1732.

\bibitem{castorena_lidar}
J.~Castorena and C.~D. Creusere, ``Sampling of time-resolved full-waveform
  {LIDAR} signals at sub-{N}yquist rates,'' \emph{IEEE Trans. Geoscience and
  Remote Sensing}, vol.~53, no.~7, pp. 3791--3802, Jul. 2015.

\bibitem{blu_oct}
T.~Blu, H.~Bay, and M.~Unser, ``A new high-resolution processing method for the
  deconvolution of optical coherence tomography signals,'' in \emph{Proc. First
  {IEEE} Int. Symposium on Biomedical Imaging: {M}acro to Nano}, vol. {III},
  Jul. 2002, pp. 777--780.

\bibitem{fri_strang}
P.~L. Dragotti, M.~Vetterli, and T.~Blu, ``Sampling moments and reconstructing
  signals of finite rate of innovation: Shannon meets {S}trang-{F}ix,''
  \emph{IEEE Trans. Signal Process.}, vol.~55, no.~5, pp. 1741--1757, May 2007.

\bibitem{prony}
G.~R. DeProny, ``Essai experimental et analytique: {S}ur les lois de la
  dilatabilit{\'e} de fluides {\'e}lastiques et sur celles de la force
  expansive de la vapeur de l'eau et de la vapeur de l'alcool, {\`a}
  diff{\'e}rentes temp{\'e}ratures,'' \emph{J. de l'Ecole polytechnique},
  vol.~1, no.~2, pp. 24--76, 1795.

\bibitem{plonka}
G.~Plonka and M.~Tasche, ``Prony methods for recovery of structured function,''
  \emph{GAMM-Mitt}, vol.~37, no.~2, pp. 239--258, 2014.

\bibitem{schmidt_music}
R.~O. Schmidt, ``Multiple emitter location and signal parameter estimation,''
  \emph{IEEE Trans. Antennas and Propag.}, vol.~34, no.~3, pp. 276--280, Mar.
  1986.

\bibitem{barabell_rmusic}
A.~Barabell, ``Improving the resolution performance of eigenstructure-based
  direction-finding algorithms,'' in \emph{Proc. IEEE Int. Conf. Acoust.,
  Speech and Signal Process. (ICASSP)}, vol.~8, 1983, pp. 336--339.

\bibitem{bdrao_rmusic}
B.~D. Rao and K.~V.~S. Hari, ``Performance analysis of root-{MUSIC},''
  \emph{IEEE Trans. Acoust., Speech and Signal Process.}, vol.~37, no.~12, pp.
  1939--1949, 1989.

\bibitem{sarkar_mp}
Y.~Hua and T.~K. Sarkar, ``Matrix pencil method for estimating parameters of
  exponentially damped/undamped sinusoids in noise,'' \emph{IEEE Trans.
  Acoust., Speech and Signal Process.}, vol.~38, no.~5, pp. 814--824, May 1990.

\bibitem{paulraj_esprit}
A.~Paulraj, R.~Roy, and T.~Kailath, ``A subspace rotation approach to signal
  parameter estimation,'' \emph{Proc. IEEE}, vol.~74, no.~7, pp. 1044--1046,
  1986.

\bibitem{eldar_2015sampling}
Y.~C. Eldar, \emph{Sampling Theory: Beyond Bandlimited Systems}.\hskip 1em plus
  0.5em minus 0.4em\relax Cambridge University Press, 2015.

\end{thebibliography}











\end{document}